\documentclass[journal]{IEEEtran}
\usepackage{amsmath,amsfonts}
\usepackage{amssymb}
\usepackage{algorithmic}
\usepackage{algorithm}
\usepackage{array}
\usepackage[caption=false,font=normalsize,labelfont=sf,textfont=sf]{subfig}
\usepackage{cite}
\usepackage{float}
\usepackage{graphicx}
\usepackage{lipsum}
\usepackage{makecell}
\usepackage{multirow}
\usepackage{textcomp}
\usepackage{stfloats}
\usepackage{url}
\usepackage{verbatim}

\usepackage{tikz,color}
\usepackage[implicit=false]{hyperref}
\hypersetup{hidelinks,
	colorlinks=true,
	allcolors=black,
	pdfstartview=Fit,
	breaklinks=true}
\definecolor{lime}{HTML}{A6CE39}
\DeclareRobustCommand{\orcidicon}{
\begin{tikzpicture}
\draw[lime, fill=lime] (0,0)
circle[radius=0.16]
node[white]{{\fontfamily{qag}\selectfont \tiny \.{I}D}};
\end{tikzpicture}
\hspace{-2mm}
}
\foreach \x in {A, ..., Z}{%
\expandafter\xdef\csname orcid\x\endcsname{\noexpand\href{https://orcid.org/\csname orcidauthor\x\endcsname}{\noexpand\orcidicon}}
}

\newtheorem{definition}{\textbf{Definition}}

\UseRawInputEncoding
\begin{document}

\title{\textsc{StatGraph}: Effective In-vehicle Intrusion Detection via Multi-view Statistical Graph Learning}
\author{
Kai Wang$^*$\hspace{-1.5mm}\orcidB{}, \IEEEmembership{Member,~IEEE}, Qiguang Jiang\hspace{-1.5mm}\orcidA{}, 
Bailing Wang$^*$\hspace{-1.5mm}\orcidC{}, \IEEEmembership{Member,~IEEE}, 

Yulei Wu\hspace{-1.5mm}\orcidD{}, \IEEEmembership{Senior Member,~IEEE}, 
Hongke Zhang\hspace{-1.5mm}\orcidE{}, \IEEEmembership{Fellow, IEEE}

\thanks{This work is supported by National Natural Science Foundation of China (NSFC) (grant number 62272129) and TaiShan Scholars (grant number tsqn202408112) .}
\thanks{Kai Wang, Qiguang Jiang and Bailing Wang are with the School of Computer Science and Technology, Harbin Institute of Technology, Weihai, China. (e-mail: dr.wangkai@hit.edu.cn; jiangqiguang\_971@163.com; wbl@hit.edu.cn)}
\thanks{Yulei Wu is with Faculty of Engineering and the Bristol Digital Futures Institute, University of Bristol, UK. (email: y.l.wu@bristol.ac.uk) }
\thanks{Hongke Zhang is with the School of Electronic and Information Engineering, Beijing Jiaotong University, Beijing 100044, China. (e-mail: hkzhang@bjtu.edu.cn)}
\thanks{Kai Wang and Bailing Wang are the corresponding authors.}
}

\markboth{SUBMIT TO IEEE Transactions}
{Shell \MakeLowercase{\textit{et al.}}: A Sample Article Using IEEEtran.cls for IEEE Journals}

\IEEEpubid{0000--0000/00\$00.00~\copyright~2025 IEEE}

\maketitle

\begin{abstract} 
In-vehicle network (IVN) is facing complex external cyber-attacks, especially the emerging masquerade attacks with extremely high difficulty of detection while serious damaging effects. 
In this paper, we propose the \textsc{StatGraph}, which is an effective and fine-grained intrusion detection methodology for IVN security services via multi-view statistical graph learning on in-vehicle controller area network (CAN) messages with insight into their variations in periodicity, payload and signal combinations. Specifically, \textsc{StatGraph} generates two statistical graphs, timing correlation graph (TCG) and coupling relationship graph (CRG), in every CAN message detection window, where edge attributes in TCGs represent temporal correlation between different message IDs while edge attributes in CRGs denote the neighbour relationship and contextual similarity. Besides, a lightweight shallow layered graph convolution network is trained based on graph property of TCGs and CRGs, which learns the universal laws of various patterns more effectively and further enhance the performance of detection. To address the problem of insufficient attack types in previous intrusion detection, we select two real in-vehicle CAN datasets covering five new instances of sophisticated and stealthy masquerade attacks that are never investigated before. Experimental result shows \textsc{StatGraph} improves both detection granularity and detection performance over state-of-the-art intrusion detection methods. Code is available at \url{https://github.com/wangkai-tech23/StatGraph}.

\end{abstract}

\begin{IEEEkeywords}
Intrusion Detection, Masquerade Attacks, Internet of Vehicles, In-vehicle Network, Controller Area Network, Multi-view Statistical Graph.
\end{IEEEkeywords}

\section{Introduction}
\IEEEPARstart{A}{s} an important part of vehicle-to-everything (V2X), intelligent connected vehicles (ICVs) equipped with in-vehicle communication networks and various embedded computing devices have begun to spread rapidly and become a popular trend \cite{beijing1}.
Along with increasing number of intelligent technologies integrated into vehicles, ICVs can obtain various intelligent services and comfortable experiences, including autonomous driving, collision avoidance, auto parking assist, and so on.
However, these technologies have also exposed the in-vehicle network (IVN) to attackers with more opportunities, and a large number of evolving new types of attacks have appeared in cyberspace, which threatens the security of IVN and even the safety of passengers \cite{beijing2invehcile}.

In terms of in-vehicle communication channels, Controller Area Network (CAN) technology is considered the de facto in-vehicle communication standard for Electronic Control Units (ECUs) in IVN embedded systems \cite{beijingmehod1}.
It can cause serious consequences if the malicious message involves the brake system, the powertrain, or the safety controls of the target vehicle \cite{beijing3attack}. With this in mind, recently researchers have chosen many methods struggle to provide security of IVNs by detecting intrusion on CAN bus \cite{Canova}, \cite{TOW-IDS}, \cite{CANShield}.
Encryption and authentication methods are effective for protecting IVNs \cite{jiami}, however, they all inevitably need to modify the existing IVN CAN protocol field definition and interaction patterns, 
which is impractical and has unacceptable high-cost since CAN protocol has been widely deployed in the production process of the automobile manufacturer around the world. 
As another security way, intrusion detection methods are usually based on bypass monitoring mechanism, without any change to the details of the existing CAN protocol or existing ECU devices, thus have great feasibility and practical advantages during actual deployment \cite{beijing4surveyIDS}.

\IEEEpubidadjcol
As artificial intelligence advances in network traffic identification, IVN intrusion detection has shifted from monitoring the physical characteristics of ECUs to analyzing CAN data flow characteristics, which reflect interactions between ECUs. Mainstream approaches predominantly use deep neural networks or graph-based methods, such as Convolutional Neural Networks (CNNs), Recurrent Neural Networks (RNNs), and graph-based intrusion detection systems \cite{wangzi2aiai}, \cite{5-20}, \cite{G-IDCS}. 
However, existing supervised neural network models label all CAN messages within a data flow as anomalous, regardless of the number of actual abnormal messages \cite{CarHackingDataset1}. This leads to coarse-grained detection, where only the overall anomaly of a detection window is identified, without fine-grained recognition of individual messages \cite{supple1-58}, \cite{supple1-32}. As a result, detection the traceability is lost \cite{CANet}, \cite{supple1-71}, \cite{reviewer1-graph}. 
Additionally, current graph-based methods primarily focus on simpler attack types \cite{wangzi2aiai}, \cite{ROADDataset}, leaving the detection of sophisticated masquerade attacks largely unexplored \cite{beijing4surveyIDS}, \cite{supple1-41}, \cite{X-CANIDS}, \cite{AERO}. 
The masquerade attacks are particularly stealthy, as they replace legitimate messages while preserving transmission rhythm and interactive behavior, but with destructive payloads. However, mainstream single-view graph-based intrusion detection methods typically focus on the relationship between CAN message IDs and frequencies. While effective, they may fail to detect masquerade attacks that alter only the payload, as they inadequately capture the coupling structure of data flow \cite{5-20}.

To address the aforementioned issues, we propose \textsc{StatGraph}, an effective in-vehicle intrusion detection approach that achieves fine-grained classification through multi-view statistical graph learning. \textsc{StatGraph} constructs node feature vectors and adjacency matrices for the Graph Convolutional Network (GCN) by generating two statistical views: the Timing Correlation Graph (TCG) and the Coupling Relationship Graph (CRG). 
First, to capture long-term statistical distributions between neighboring messages, \textsc{StatGraph} generates global node features through statistical attributes in the TCG. In one TCG, edges represent the temporal correlations between different CAN IDs within detection windows. By incorporating these temporal correlations, \textsc{StatGraph} merges global features from TCGs with local features from the data content, enabling more accurate intrusion detection. Second, the edge attributes in the CRGs not only capture the neighbor relationships but also reflect similar properties of the same CAN IDs over extended time periods, enhancing the contextual associations within detection windows. Additionally, the GCN's suitability for small-scale graphs in streaming scenarios, coupled with its higher computational efficiency, makes it ideal for IVN intrusion detection, where fast detection is critical.

The main contributions of this paper can be summarized as follows:

\begin{itemize}
    \item {We propose \textsc{StatGraph} to effectively detect IVN attacks enhancing detection by adopting multi-view statistical graph learning, which considers both the interactive relationships of CAN messages and their payload-level contextual similarities. This approach enables \textsc{StatGraph} to accurately identify variations in periodicity, payload, and signal combinations, thereby detecting a wide range of attacks, including sophisticated and sophisticated masquerades.}
    \item {Current deep learning-based approaches build the model input with multiple CAN messages, resulting in coarse recognition and the inability to locate the malicious message. To accurately quantify the granularity level of detection, we introduce a new criterion for granularity evaluation, Identification Granularity (IG), which is used to depict the recognition accuracy for each message at a fine-grained granularity rather than coarse-grained on a whole window of messages.}
    \item {To evaluate the classification ability of \textsc{StatGraph} on captured CAN messages of realistic in-vehicle environment, we conduct experiments on an in-vehicle computing platform (NVIDIA Jetson Nano T206 with ARM architecture) and on a personal computer (LENOVO 90VA000JCP), respectively. Besides, we select the Car Hacking Dataset and the ROAD Dataset to cover as many types of attacks as possible, which are both directly generated from real vehicles on sale. Notably, to the best of our knowledge, we are the first to investigate 5 new types of masquerade attacks introduced by the ROAD Dataset, which can cause incorrect working status of physical components within real vehicles (e.g., wrong speed indicator position by injecting fake wheel speed values) and bring in severe damage on both vehicles and passengers. 
    }
\end{itemize}


The reminder of this paper is organized as follows: Section \ref{section3} introduces the threat models and summarizes existing representative IVN intrusion detection methods. Section \ref{section4} describes our proposed \textsc{StatGraph}. We show the experiment setup and results in section \ref{section5} followed by the discussion in section \ref{section6}. We conclude the paper in section \ref{section7}.

\section{Background}\label{section3} 

\subsection{Threat Model}
ECUs use bit-by-bit arbitration and explicit broadcasting to deliver CAN protocol messages, enabling fast and efficient data exchange. An adversary can access the CAN bus via the OBD-II port or exploit the Original Equipment Manufacturer (OEM)’s telematics service through remote external network access. Depending on the adversary’s objectives, attacks are typically categorized into three types.

\textit{1) Fabrication Attacks}: 
An adversary compromises an ECU and injects malicious messages with forged IDs and data fields into the CAN bus. This is a common and straightforward attack, easily executed. Despite this, all legitimate ECUs remain active, transmitting their original data. Fabrication attacks increase CAN bus traffic, leading to dynamic and distributed changes in communication behavior. Examples include Denial of Service (DoS) attacks, fuzzy attacks, and targeted ID attacks.

\begin{figure*}[htbp]
	\centering
	\includegraphics[width=1.0\linewidth]{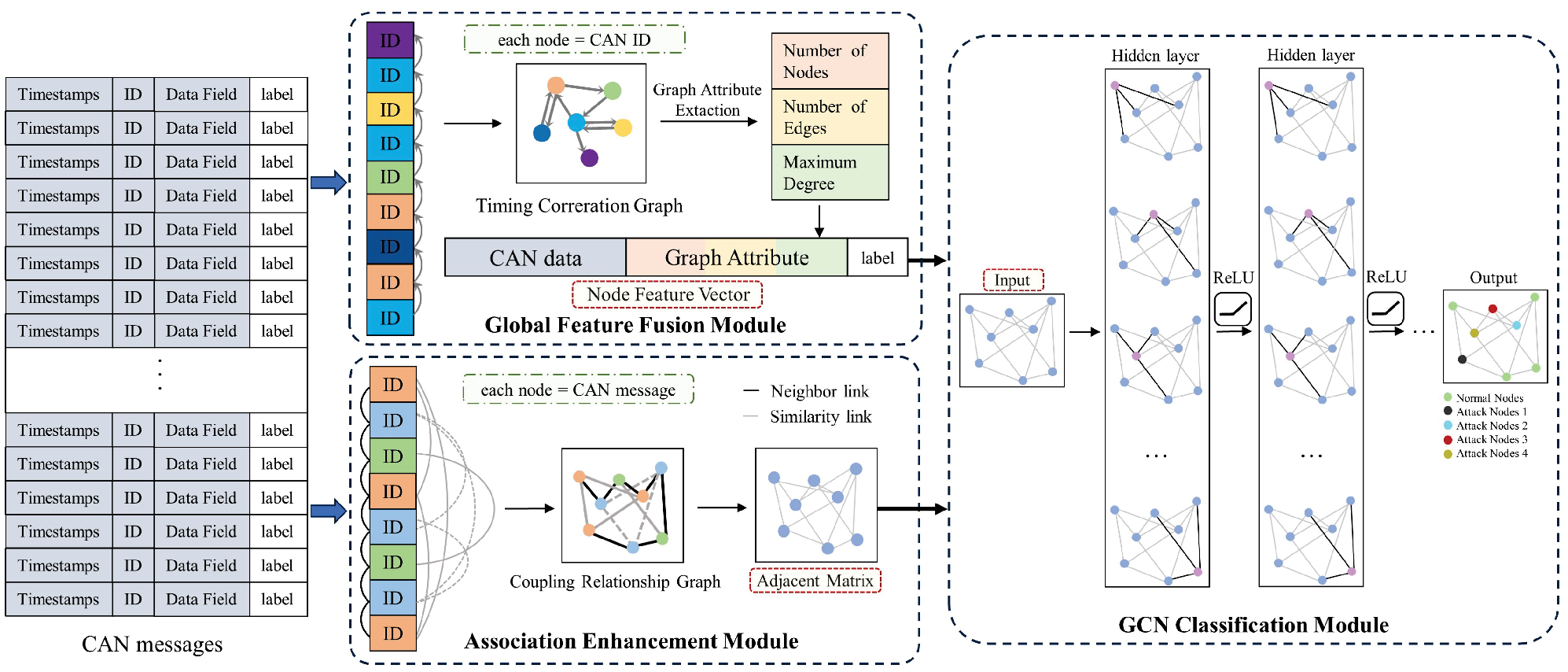}
	\caption{The architecture of \textsc{StatGraph}. Global Feature Fusion Module (\ref{section4.2.1}) generates \textit{node feature vectors} by constructing TCGs that react to the long-term distribution of IVN CAN messages. Association enhancement module (\ref{section4.2.2}) produces \textit{adjacency matrixes} by constructing CRGs whose edges reflect the topological relationship information between adjacent CAN messages or those with the same ID. GCN classification module (\ref{section4.2.3}) accepts the generation of streaming graph-structured data from the global feature fusion module and association enhancement module to perform training and detection tasks.}
	\label{overview}
\end{figure*}

\textit{2) Suspension Attacks}: In this type of attacks, legitimate ECUs can be disabled by an adversary. Specifically, the targeted ECU is weakly compromised and no longer able to send messages, affecting the ability of other ECUs to use some of the information from the targeted ECU to function properly. 
As a result, suspension attacks can cause damage to not only weakly compromised ECU itself, but also other receiver ECUs.

\textit{3) Masquerade Attacks}: They are a combination of fabrication and suspension attacks, which keeps the most sophisticated and stealthy attack behavior patterns.
The adversary first performs a suspension attack on the ECU that he wants to replace, making it disappear from the CAN network. Afterwards, another fully compromised ECU is used to send spoofed messages with a given ID at a realistic frequency, thus masquerading as an unavailable target ECU. 
Using this more advanced attack strategy, only the payload of involved CAN messages are stealthily changed, but the original behavioural characteristics of the target ECU can be completely maintained without generating message conflicts or causing changes in the dynamics and distribution of the system communication behaviours, which causes a large number of intrusion detection methods to fail. 

\subsection{Related Work}\label{SectionRelatedWork}


Deep learning-based intrusion detection methods have great potential in securing IVNs, yet significant shortcomings in terms of detection granularity and sophisticated attack detection, which limits their application in IVNs \cite{beijing3attack}, \cite{TOW-IDS}, \cite{beijing4surveyIDS}, \cite{X-CANIDS}, \cite{AERO}.
On the one hand, coarse-grained detection can only identify whether an entire CAN message detection window is abnormal rather than fine-grained identification on every single message \cite{supple1-58}, \cite{supple1-32}), which reduces the reliability of this type of approaches.
On the other hand, with the continuous evolution of attack variants, there emerges more advanced masquerade attacks whose payload content is destructive but the sending rhythm can stealthily replace the sending behavior of legitimate messages, resulting in the inability of existing methods to cope with such attacks \cite{Canova}, \cite{CANShield}, \cite{ROADDataset}. 

Besides, graph-based IVN intrusion detection methods have emerged due to their powerful abstraction capability on learning the characteristics of data interaction between heterogeneous ECUs to reflect the collaborative process and data distribution pattern of CAN bus. 
However, the detection of relatively simple types of IVN attacks is overly investigated in current graph-based methods \cite{5-20}, while the research efforts on effectively detecting sophisticated and stealthy masquerade attacks are still quite lacking. 
The studies \cite{5-20}, \cite{G-IDCS}, \cite{Federated-Graph}, \cite{graph2} only use the ID of CAN data, ignoring the necessity of learning the pattern of payload variations, which makes them impotent in detecting masquerade attacks that completely replicate the statistical behavior patterns of legitimate messages and with only the payload thresholds being altered. 
Moreover, current graph-based IVN intrusion detection methods are mainly coarse-grained (e.g., the injection attack is detected once every 1200 CAN messages in \cite{5-20} and once every 200 CAN messages in \cite{G-IDCS}), which clearly fails to meet the real-time requirements of IVN security. 
In addition, different from the most existing methods using single view learning paradigm \cite{5-20}, \cite{G-IDCS}, \cite{Federated-Graph}, \cite{graph2} on IVN intrusion detection, \textsc{StatGraph} starts exploring the utilization of multi-views to represent the real graph data, where different graph views express different kinds of relationships between nodes.

\section{Methodology}\label{section4}
In this section, we give the system overview of the \textsc{StatGraph}, and then present the details of its key components. 

\subsection{System Overview}

\textsc{StatGraph} consists of the following three components, as illustrated in Figure \ref{overview}.

%

$\bullet$ \textbf{Global Feature Fusion Module} (\ref{section4.2.1}) generates \textit{node feature vectors} by constructing TCGs that capture long-term distribution of CAN messages. Each distinct CAN ID within a detection window is treated as a node in the TCG. The statistical properties of these nodes are global, meaning that all CAN messages with the same ID, regardless of when they are generated, are aggregated into the same node in the TCG.

$\bullet$ \textbf{Association Enhancement Module} (\ref{section4.2.2}) constructs CRGs to produce \textit{adjacency matrices}, where edges reflect the topological relationships between adjacent CAN messages or those with the same ID. This approach enhances the coupling structure compared to traditional adjacency matrices. In CRGs, each CAN message within a detection window is treated as a node, and an edge is placed between two adjacent messages. Additionally, two CAN messages with the same ID are represented as separate nodes, with an edge indicating their topological relationship.

$\bullet$ \textbf{GCN Classification Module} (\ref{section4.2.3}) processes the graph-structured data generated by the Global Feature Fusion and Association Enhancement modules to perform training and inference tasks. The graph data used by the GCN model integrates the complex communication patterns and dynamic changes of IVN CAN messages, as learned from the TCGs and CRGs. This integration significantly enhances the training quality and classification performance of the GCN model.
 
%

We define two different types of generated multi-view. 
\begin{definition}\label{def3}
We consider two types of graphs, i.e., TCG $G(V,E)$ for long-term statistical attribute and CRG $\hat{G}(\hat{V},\hat{E})$ for short-term messages topological structure, where $v_i,v_j\in V$ and $\hat{v}_i,\hat{v}_j\in \hat{V}$ denote the nodes; $ e_{ij}\in E\subseteq V\times V$ and $\hat{ e}_{ij}\in \hat{E}\subseteq\hat{V}\times\hat{V}$ represent the edges. Here, $w_{ij},~\hat{w}_{ij}$ are the weight of edge $e_{ij}$ and $\hat{e}_{ij}$ respectively.
\end{definition}
\subsection{Global Feature Fusion Module}\label{section4.2.1}

In this module, we design the TCG representing the CAN distribution over long-term to extract the interaction features of CAN data stream sequences.


$\bullet~${\bf{Generation of TCG based on CAN stream}} 

The CAN flow can be regarded as multivariate time series.
\begin{definition}\label{def1}(Multivariate time series)
CAN data is a multivariate time serie $\Psi \in \mathbb{R}^{T \times V}$:
\begin{equation}\label{defCANdata}
  \Psi = \{\psi_1,\psi_2,\ldots,\psi_t,\ldots,\psi_T\},
\end{equation}
where $ \psi_t = \{timestamp,ID,DLC,d_1,\ldots,d_8,\mathrm{type}\}\in \mathbb{R}^{V}$. $V=12 $ is the dimension of a recorded data CAN message.

We choose $ID = d_0$, data field $ \{d_1,d_2,\ldots,d_8 \}\in \mathbb{R}^{8}$ and label value $d_l$ to form $\phi_t = \{d_0,d_1,d_2,\ldots,d_8,d_l \}\in\mathbb{R}^{10}$ as inputs representation for \textsc{StatGraph}. 
\end{definition}

%
%
%

Since the recurrent patterns of IDs are one of the manifestations of the periodicity for CAN messages, we divide IVN CAN messages into numerous detection windows and characterize the message distribution under each detection window separately to guarantee the CAN data pre-processing of the \textsc{StatGraph} in a more efficient way.

We divide the CAN detection windows in the following:
\begin{definition}\label{def2}
A series of CAN messages monitored from the IVN arranged according to the normal communication sequence can be regarded as located within a single detection window. In proposed algorithm, we consider $N$ messages to be a detection window \cite{5-20}. $N$ is the detection window size. Let's assume that there are $K$ detection windows in total. $w_k$ denotes the matrix consisting of CAN messages within $k$-th detection window after preprocessing.
\begin{equation}\label{defwin}
  w_k = \{\phi _{k*N+1},\phi _{k*N+2},\ldots,\phi_{k*N+N} \}.
\end{equation}
Therein, $k \in \{0,1,2,\ldots,K-1\}$.
\end{definition}
 
After dividing the CAN messages into $K$ detection windows according to Definition \ref{def2}, our scheme designs Algorithm \ref{algorithm1} to track legitimate transitions between the IDs of all two consecutive CAN messages and finally establish a TCG list, through multivariate time series CAN messages. Specifically, for each set of $N$ messages within a detection window, Algorithm \ref{algorithm1} treats each distinct CAN ID as a node and constructs edges to represent ID-to-ID relationships. For clarity, $Set[ID_i] =  \{ \phi_n | \phi_n[d_0] = ID_i,~\phi_n\in w_k\}$ denotes the set of messages sharing the same $ID_i$, where $ID_i$  serves as the key for $Set[ID_i]$.The algorithm iterates over all detection windows (Line 2). For the $k$-th detection window, it collects each unique ID and creates a new key (ID)-value (message) pair upon the first occurrence of the ID (Lines 4 and 8). Subsequently, an edge is established between the IDs (nodes) corresponding to two consecutive messages. If the edge already exists, its weight is incremented (Line 11).

\begin{equation*}
w_{i,j}\leftarrow 
\left\{
\begin{aligned}
& +1,~\phi_n \in Set[ID_j],~\phi_{n+1} \in Set[ID_i],\\
& +0 ,~ else.\\
\end{aligned}
\right.
\end{equation*}
The algorithm keeps the graph $G_k$ built by the current detection window (Line 14) and eventually returns the graphs ${G_1, \ldots, G_K}$ for all detection windows.

\begin{algorithm}[htb]
	\caption{TCG Building Algorithm.}
	\label{algorithm1}
	\begin{algorithmic}[1] 
		\REQUIRE ~~\\
		Detection Window List of CAN Messages: $CANWindowList$ = $\{w_1,w_2,\ldots,w_K\}$, therein $w_k $= $\{\phi _{k*N},\phi _{k*N+1},\phi _{k*N+2},\ldots,\phi_{k*N+N} \}.$ 
		\ENSURE ~~\\ 
		$CorrelationGraphList[G_1,G_2,\ldots,G_K]$ $\rhd$ Timing Correlation Graph array of CAN bus data;
        \STATE $CorrelationGraphList\leftarrow[~],~IDDictionary\leftarrow\{"~":~\}$
        \FOR {$w_k$ in $CANWindowList$}
            \STATE Initialize $Graph$; \\
            \STATE $LastID \leftarrow \phi _{k*N}[d_0]$; \\
            \STATE $IDDictionary[LastID] \leftarrow len(IDDictionary)$ $\rhd$ Create a new key/value pair in the dictionary; \\
            \FOR {$i$ in $N$}
                \STATE $NowID \leftarrow \phi _{k*N+i}[d_0]$; \\
                \IF {not NowID in IDDictionary.keys()}
				    \STATE $IDDictionary[NowID] \leftarrow len(IDDictionary)$ $\rhd$ Create a new key/value pair in the dictionary;\\
		        \ENDIF
                \STATE Connect an edge from $v_{IDDictionary[NowID]}$ to $v_{IDDictionary[LastID]}$ $\rhd$ Create link between two graph nodes;\\
                \STATE $LastID \leftarrow NowID$; \\

            \ENDFOR
            \STATE Append $Graph$ to the $CorrelationGraphList$;\\
            \STATE $IDDictionary.clear$ $\rhd$ clear $IDDictionary$

        \ENDFOR
	\end{algorithmic}
\end{algorithm}

$\bullet~${\bf{Building~feature~vector~of~each~node}}

After construction of TCG, graph feature vectors can be generated by statistical attributes of TCG, such as: the number of edges, the number of nodes, the max weight, the average weight, the max degree, etc. Our proposed method characterizes feature vector for each CAN message, which could be defined as $X_{k,n}$, denoting the node feature of $n$-th message in $k$-th detection window of raw CAN messages, $n\in \{1,2,\ldots,N\}.$ Hence, denote node feature matrixes of $k$-th detection window as:
\begin{equation}\label{X_k}
  X_k,~k \in \{0,1,\ldots,K-1\}
\end{equation}
 

According to research \cite{5-20}, we choose extracting the node number, edge number, maximum degree from single constructed TCG and consider them as global features which will be supplemented to node feature vectors.

It should be noted that sophisticated masquerade attacks may alter the payload field of CAN messages instead of the ID area. Therefore, \textsc{StatGraph} must be capable of recognizing changes in the payload patterns of each CAN message. Additionally, detecting masquerade attacks could be more precise by merging the localized payload information of each CAN message with the global characteristics of the detection window to enhance the anomaly detection capability.
Hence, node feature $X_{k,n}=\{d_0,d_1,d_2,\ldots,d_8,x_9,\ldots,x_{11} \}\in~ \mathbb{R}^{12}$, where $d_0$ is CAN ID converted to decimal, $d_1,d_2,\ldots,d_8$ means data value in data field and $x_9,x_{10},x_{11}$ represent node number, edge number, maximum degree of the TCG to which the node belongs, respectively. 
To ensure detection granularity for each CAN message, the node feature $X_{k,n}$ retains the label indicating whether it is normal or injected, though it is not part of the input node feature vector.

Additionally, simpler attacks can be efficiently detected by leveraging the statistical properties of TCGs, which capture the long-term distribution of in-vehicle communication dynamics. For example, when an attacker injects a large volume of messages or alters the frequency of CAN messages from certain ECUs, it can significantly disrupt the probability distribution and payload signaling patterns of the CAN messages.


\subsection{Association Enhancement Module}\label{section4.2.2}
This module leverages CRGs to enhance the association between CAN messages that exhibit similarity and short-term dependence. While TCGs capture periodicity and regularity, they cannot detect masquerade attacks, which alter payloads without affecting statistical patterns. Legitimate message payloads fluctuate but exhibit regularities within the same ECU \cite{shen[4]}, and CAN messages from different ECUs may share common patterns, which masquerade attacks disrupt. 

CRGs capture these patterns, or coupling relationships, by reinforcing connections between messages, emphasizing their short-term correlations. Unlike other methods focused solely on contextual connections due to the simplicity of the CAN format, CRGs combine both contextual correlations and content similarity, offering a more comprehensive understanding of CAN message associations.

Despite significant efforts to model graph structural information, existing techniques are limited in their ability to capture coupled relationships \cite{CarHackingDataset1}. 
In \textsc{StatGraph}, the coupling characteristic of the data flow is defined as the contextual correlation between neighboring CAN messages and the similarity of messages with the same ID. This results in a topological relationship-aware messaging paradigm, reflecting the coupling relationship. A novel method is introduced to construct the adjacency matrix by creating Coupling Relationship Graphs (CRGs), as shown in Algorithm \ref{algorithm2}.

%
%


\begin{algorithm}[htb]
	\caption{CRG Building Algorithm.}
	\label{algorithm2}
	\begin{algorithmic}[1] 
		\REQUIRE ~~\\%
		Detection Window List of CAN Messages: $CANWindowList$ = $\{w_1,w_2,\ldots,w_K\}$, therein $w_k $= $\{\phi _{k*N},\phi _{k*N+1},\phi _{k*N+2},\ldots,\phi_{k*N+N} \}.$
		\ENSURE ~~\\ %
        $RelationshipGraphList[\hat{G}_1,\hat{G}_2,\ldots,\hat{G}_K]$ $\rhd$ Coupling Relationship Graph array of CAN bus data;		
        \STATE $RelationshipGraphList\leftarrow[~],~IDSet\leftarrow\{"~":~\}$
        \FOR {$w_k$ in $CANWindowList$}
            \STATE Initialize $Graph$; \\
            \STATE $LastID \leftarrow \phi _{k*N}[d_0]$; \\
            \STATE $IDSet[LastID] \leftarrow len(IDSet)$ $\rhd$ Create a new key/value pair in the dictionary; \\
            \FOR {$i$ in $N$}
                \STATE $NowID \leftarrow \phi _{k*N+i}[d_0]$; \\
                \STATE $ (\hat{v}_{i-1}, \hat{v}_i) \leftarrow 1$  $\rhd$  Link two neighbor nodes (Contextual correlation property);\\
                \IF {not NowID in IDSet.keys()}
				    \STATE $IDSet[NowID]  \leftarrow [i]$ $\rhd$ Create a new key/value pair in the dictionary;\\
		        \ELSE
                    \STATE Append $i$ to the $IDSet[NowID]$;\\
                    \STATE Link the edges between nodes in $ IDSet[NowID]$ $\rhd$ Connect all nodes with the same ID value (Similarity);\\
                \ENDIF
		
            \ENDFOR
            \STATE Append $Graph$ to the $CorrelationGraphList$;\\
            \STATE $IDSet.clear$ $\rhd$ clear $IDSet$
		
        \ENDFOR
	\end{algorithmic}
\end{algorithm}
  
From Definition \ref{def2}, we set the same detection windows as section \ref{section4.2.1}.
And we treat each CAN message in $k$-th detection window as a node to construct CRG with $N$ nodes, whose edge represents the short-term relationship between messages.
A loop is iterated over each detection window (Line 2).
In $k$-th detection window, we define the nodes with the same ID as a set $IDSet[ID_i] =  \{ \phi_n | \phi_n[d_0] = ID_i,~\phi_n\in w_k\}$, which is constructed by counting the messages corresponding to each ID (Lines 5 and 9).
%
%
%
Next, the algorithm constructs edges of CRG $\hat{G}_k$ by searching coupling relationships.
For $i,j$ in $\{1,2,\ldots,N\}$, we have
\begin{equation}\label{gongshi1}
\hat{e}_{ij}=
\left\{
\begin{aligned}
& 1,~j=i+1~\mathrm{or}~i=j+1,\\
& 1,~\phi_j,\phi_{i} \in Set[ID_m],\\
&0 ,~ \mathrm{else}.\\
\end{aligned}
\right.
\end{equation}

The first row in equation \ref{gongshi1} specifies that an edge is created between nodes representing two consecutive CAN messages (Line 7). The second row in equation \ref{gongshi1} indicates that nodes sharing the same CAN ID are interconnected (Line 12). The last row in equation \ref{gongshi1} denotes that edges not belonging to the aforementioned relationships are excluded from connection.
 
Finally, Algorithm \ref{algorithm2} returns undirected graphs (CRGs) for multiple detection windows, whose adjacency matrixes will be used to classification task in GCN via learning the universal rule. Therefore, the adjacency matrix of $k$-th detection window as input for next GCN classification module can be defined as:
\begin{equation}\label{A_k}
  A_k,~k \in \{0,1,\ldots,K-1\}
\end{equation}
 
CRG serves as a helpful complement to learn the underlying laws of the CAN bus, which can greatly improve the ability of intrusion detection model to recognize masquerade attacks.

\subsection{GCN Classification Module}\label{section4.2.3}
We propose a multi-layer GCN model for classifying attacks on the in-vehicle CAN bus, trained with graph-structured data from the Global Feature Fusion and Association Enhancement modules that capture IVN CAN message interactions. 

The GCN model \cite{gcn_first} applies first-order graph convolution to encode both structure and node features. Node features are enhanced by global patterns from TCG statistical properties, and the adjacency matrix governs message passing, with aggregated messages weighted accordingly. CRG design is crucial to improving performance. GCN node classification distinguishes normal and injected attributes for each CAN message, emphasizing the need for labeled data.

The multi-layer GCN can be formulated as:
\begin{equation*}
  H^{(l+1)} = \sigma \left( \tilde{D}^{-1/2}\tilde{A}\tilde{D}^{-1/2} H^{(l)} W^{(l)}\right)
\end{equation*} 
where $H^{(l)}$ is the matrix of activations in the $l$-th layer, $W^{(l)}$ denotes the trainable weight matrix of $l$-th layer. $\sigma(\cdot)$ is the activation function. Denote number of hidden layer units as $h$, and $l\in \{0,1,2,\ldots,L\}$.
Set the adjacency matrix of input graph as $A$. $\tilde{A} = A + I_{M}$, where $M$ is the size of $A$ and $I_{M}$ is the $M$-dimensional identity matrix, $\tilde{D}$ is degree matrix of $ \tilde{A}$ and $\tilde{D}_{ii} = \sum_{j}\tilde{A}_{ij} $. Denoting the node feature matrix of input graph as $X$, we can get initial $H^{(0)}=X$. 
After multiple experiments, from the perspective of lightweighting and performance, our GCN model consists of an input layer, $L$ hidden layers and a dense layer, shown in Table \ref{Structure1}.

\begin{table}[htb]\scriptsize 
  \centering
  \caption{Structure and dimensions of all layers of \textsc{StatGraph}.}\label{Structure1}
  \setlength{\tabcolsep}{1.5mm}{
  \begin{tabular}{ccccccccccccc}
    \hline
    \hline
   $Role$ & Layer Name & Data Shape & Activation &  Dropout~Ratio  \\
   \hline Input & Input &  (B$ * N) \times$ 12 & - & - \\
   \hline  \multirow{10}{*}{\makecell{Feature\\ Extraction }}  & GCN 1 & (B$ * N)\times h$ &  ReLU  & -  \\
                             & Dropout 1 &(B$ * N)\times h$  & - & 0.5  \\
                             & GCN 2 & (B$ * N)\times h$  &  ReLU  & -  \\
                             & Dropout 2 & (B$ * N)\times h$  & - &  0.5  \\
                             & GCN 3 & (B$ * N)\times h$  &  ReLU  & -  \\
                             & $\cdots$ & ~  & ~ &  ~  \\
                             & Dropout $l-1$~ & (B$ * N)\times h$  & - &  0.5  \\
                             & GCN $l$~ &(B$ * N)\times h$  &  ReLU  & -  \\
                             & $\cdots$ & ~  & ~ &  ~  \\
                             & Dropout $L-1$ & (B$ * N)\times h$  & - &  0.5  \\
                             & GCN $L$ &(B$ * N)\times h$  &  ReLU  & -  \\
   \hline  \makecell{Multi-\\classification} 
                               & Dense  & (B$ * N)\times F $  & Sigmoid  & -  \\
    \hline
    \hline
  \end{tabular}}
\end{table}
In order to improve the training efficiency of the model in coding, we set $batchsize=B$, which means that we take the node feature matrixes and the adjacency matrixes corresponding to $B$ consecutive detection windows as once input. 
Hence, the input node feature matrix $X\in\mathbb{R}^{(B*N)\times12}$ is the continuous $X_k$ splice of $B$ and the input adjacency matrix $A$ with size $(B*N,B*N)$ is a diagonal concatenation of $B$ continuous matrixes $A_k$. 

To capture the fine-grained details of \textsc{StatGraph}, the GCN is designed to perform node classification, outputting abstract features for each CAN message.

%
For $k$-th detection window, the last dense layer of GCN model returns output $Z_k = \{Z_{k,1},\ldots,Z_{k,N}\} $, where $Z_{k,n}\in \left\{0,1,2,\ldots, F\right\}$ is employed to predict the attribute labels of the $n$-th message in the $k$-th detection window. 
In the training phase, the $cross-entropy$ loss function compares real label of the $n$-th message in the $k$-th detection window to $Z_{k,n}$ through a sigmoid activation function.

Besides, L2 regularization is used to restrict the values of weights to prevent overfitting. Besides, our GCN model is trained using the Adam optimizer, with a learning rate of $10^{-3}$. In addition, we deploy $dropout$ layers with a rate of 0.005 after each $GraphConv$ layer. Moreover, we set the rectified linear unit ($ReLU$), a widely used nonlinear function, as activation function in our model. It provides the neural network with nonlinear and allows neural network learning complex relationships in the data. And hyperparameters for training the \textsc{StatGraph} are shown in Table \ref{biaoparameter}.

\begin{table}[htb]\footnotesize
  \centering
  \caption{Hyperparameters for training the \textsc{StatGraph} model.}\label{biaoparameter}
  \setlength{\tabcolsep}{1.5mm}{
  \begin{tabular}{ccc}
    \hline
    \hline
   $Hyperparameters$ & Value(Car Hacking) & Value(ROAD)  \\
   \hline $B$ (Batchsize) & 40 & 5\\
    $N$ (Detection Window Size) & 50 & 400\\
    $L$ (Total Number Of Layers) & 4 & 1 \\
    $h$ (Number of Hidden Layer Units) & $32$ & $32$\\
    $F$ (Number of Categories) & 5 & 6 \\
     Learning Rate & $5\times 10^{-4}$ & $5\times 10^{-4}$\\ 
      Weight Decay & $5\times 10^{-4}$ & $5\times 10^{-4}$\\
        \hline
        \hline
  \end{tabular}}
\end{table}

This module makes full use of node features fusing statistical regulation represented by recurrent patterns of ID and adjacency matrices augmenting association relations for training, as a way to achieve high-performance fine-grained detection and classification of multiple attacks.

\section{Experiment}\label{section5}
In this section, we implemented the \textsc{StatGraph} framework under two different computing environments (see \S \ref{sec_experiment_A}), and the evaluation metrics for quantifying the detection level of granularity are presented in \S \ref{sec_experiment_B}. The typical baseline methods for horizontal comparisons are described in \S \ref{sec_experiment_C}, and the details of the Car Hacking Dataset and ROAD Dataset are given in \S \ref{sec_experiment_D}. Extensive experiments are conducted on evaluating the running performance and the detection performance in \S \ref{sec_experiment_E} and \S \ref{sec_experiment_F} respectively, and the fine-grained potential exploration is conducted in \S \ref{sec_experiment_G}. Finally, the effects of different parameters on \textsc{StatGraph} are analyzed in \S \ref{sec_experiment_H}.

\subsection{Environment}\label{sec_experiment_A}

As a core technology for vehicle network security, the intrusion detection model must offer real-time detection capabilities and a low false positive rate, particularly in vehicle environments with limited computational resources. 


With this in mind, we selects the NVIDIA Jetson Orin Nano hardware platform (based on the Xavier system-on-chip), which is widely used in the automotive industry and academia. It features a GPU, CAN bus support, and an ARM architecture, preloaded with the Ubuntu 20.04 operating system and a variety of I/O interfaces. With approximately 21 TOPS (Tera Operations Per Second) of computing power and low power consumption, the device is well-suited for running and evaluating complex intrusion detection algorithms in resource-constrained on-board environments, thereby validating the model’s feasibility and effectiveness.

\begin{figure}[!ht]
	\centering	
    \includegraphics[width=0.45\linewidth]{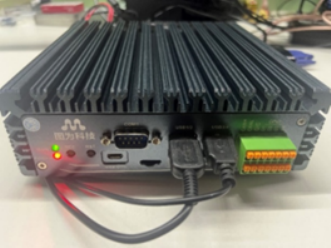}
    \includegraphics[width=0.45\linewidth]{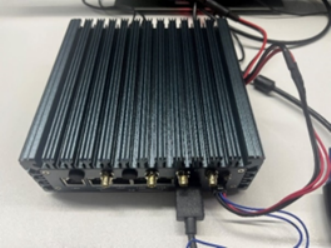}
	\caption{NVIDIA Jetson Nano T206.}
	\label{T206}
\end{figure}

The Jetson Orin Nano platform supports medium-sized deep learning model inference, meeting the computational needs for tasks such as target detection and time-series data analysis. The T206S in-vehicle computing platform from the Jetson Orin Nano series, shown in Figure \ref{T206}, features an ARMv8 Processor rev 1 and a 512-core NVIDIA Ampere architecture GPU with 8.0GB of memory. It runs on Ubuntu 20.04, using PyTorch 1.12.1 and Python 3.8.10.

For training, we utilize a more powerful device, the LENOVO 90VA000JCP, to simulate a cloud environment and accelerate the training process. It features a 64-bit Intel Core i7-13700 CPU at 3.6GHz and a Geforce RTX 4080 32GB GPU, with PyTorch 1.13 and Python 3.8 for the \textsc{StatGraph} model. In comparison, in-vehicle devices like the NVIDIA Jetson Orin NX, with 100 TOPS, provide over 70 TOPS of computational power, while the device used here delivers 49 TOPS, sufficient for model validation.


\subsection{Evaluation Metrics}\label{sec_experiment_B}
To evaluate the effectiveness of different methods, this paper used classic classification metric method.
Accuracy, Precision and Recall can be calculated by using the number of true naegatives (TN), true positives (TP), false negatives (TN) and false positives (FP), defined as follow:
\begin{equation}\label{acc}
    \mathrm{Accuray} = \frac{TP+TN}{TP+TN+FP+FN}
\end{equation}
\begin{equation}\label{pre}
    \mathrm{Precision} = \frac{TP}{TP+FP} 
\end{equation}
\begin{equation}\label{recall}
    \mathrm{Recall} =  \frac{TP}{TP+FN}  
\end{equation}
Besides, F1-score is more useful than accuracy in the case of unbalanced class distribution. It is the harmonic mean of precision and recall, calculated as follow:
\begin{equation}\label{f1}
    \mathrm{F1-score} =  \frac{2 \times \mathrm{Precision}\times \mathrm{Recall}}{\mathrm{Precision+Recall}} 
\end{equation}


IG is designed to evaluate the identification accuracy of intrusion detection methods at a fine-grained level.
Based on the correct identification of the detection window (that is, the label of the $k$-th detection window, $label_{k}$ = the predicted value of the $k$-th detection window, $predict_{k}$), IG is calculated through TI (True Identification), which represents the number of correctly identified samples per detection window. Hence, The IG is defined as follows:

\begin{equation}\label{IG}
   \mathrm{IG} =  \frac{1}{N}\sum_{k=1}^{K}\mathrm{TI}_k\times C_k, 
\end{equation}
where
\begin{equation*}
C_k =  
\left\{
\begin{aligned}
& 1,~ label_{k} = predict_{k},\\
& 0,~ label_{k} \neq predict_{k}.\\
\end{aligned}
\right.
\end{equation*}

\subsection{Compared Methods}\label{sec_experiment_C}
To ensure a fair comparison, several representative and advanced models are selected for binary and multi-classification tasks, as well as for coarse-grained and fine-grained levels:
\begin{itemize}
    \item \textit{Graph-based intrusion detection system (Graph-based IDS)} \cite{5-20}: A coarse-grained, graph-based IVN intrusion detection method that uses the median and Chi-squared tests for binary classification tasks. It operates on the number of nodes, edges, and maximum degree of graph-structured data generated from CAN messages over six detection windows.
           
    \item \textit{G-IDCS TH\_classifier} \cite{G-IDCS}: Similar to Graph-based IDS, this graph-based IVN intrusion detection method performs binary classification based on thresholds for the number of nodes, edges, and maximum degree of graph-structured data within each detection window.
    
    \item \textit{G-IDCS ML\_classifier} \cite{G-IDCS}: Using the same graph generation process as G-IDCS TH\_classifier, this method performs multi-class classification tasks using a Random Forest (RF) algorithm on the generated graph-structured data within each detection window.
 
    \item \textit{EfficientNet} \cite{EfficientNet}: A neural network architecture with a novel compound scaling method that balances network width, depth, and resolution. It performs coarse-grained multi-classification intrusion detection tasks in resource-constrained IVN environments.
    
    \item \textit{MobileNetV3} \cite{MobileNetV3}: A lightweight neural network that utilizes efficient depthwise separable convolutions, suitable for IVN multi-classification intrusion detection tasks at a coarse-grained level (window-level classification).
    
    \item \textit{CANet} \cite{CANet}: The first deep learning-based method for IVN multi-classification intrusion detection, which checks individual CAN messages using a tailored network structure that fits the signal space of CAN data for each detection window.
 
    \item \textit{CAN-RF} \cite{114}: A fine-grained multi-classification method based on an integrated learning framework where each decision tree is constructed from a random subset of features of each CAN message.
    
    \item \textit{MultiLayer Perceptron (CAN-MLP)} \cite{114}: This model executes multi-class classification tasks on each CAN message using a feedforward neural network structure.
     
    \item \textit{Long Short Term Memory (CAN-LSTM)} \cite{20}: A popular RNN variant that predicts the next CAN ID more effectively based on previous observations, making it suitable for fine-grained multi-classification tasks.
\end{itemize}

\subsection{Datasets Details}\label{sec_experiment_D}
To the best of our knowledge, seven publicly available CAN datasets with labeled attacks exist (see Table \ref{biaoOpenDatasets}). We prioritize datasets with real attacks on actual vehicles, as simulated data may lack real-world fidelity. The Car Hacking Dataset, commonly used in IVN intrusion detection research, contains multiple fabrication attacks. In contrast, the ROAD Dataset, known for its high fidelity, includes diverse masquerade attacks and validates their physical effects in real vehicles. For instance, injecting fake wheel speed values can trigger critical malfunctions, posing risks to both the vehicle and passengers. While prior IDS research has focused on fabrication attacks, masquerade attacks remain underexplored. Therefore, this study selects the Car Hacking and ROAD Datasets to evaluate the effectiveness of our detection methods. Detailed descriptions and experimental setups for each dataset are provided in the following sections.

%

\subsubsection{Car~Hacking~Dataset}\label{subsec_CarHacking}

%
%

\begin{table}[htb]\scriptsize
  \centering
  \caption{Open CAN Datasets.}\label{biaoOpenDatasets}
  \setlength{\tabcolsep}{1.4mm}{
  \begin{tabular}{cccc} 
    \hline
     \multirow{2}*{Dataset} & \multirow{2}*{\makecell{ Real/\\Synthetic}} &\multirow{2}*{\makecell{ Fabrication \\ Attack}}  & \multirow{2}*{\makecell{ Masquerade \\ Attack}} \\
   ~  & ~  & ~ & ~ \\
   \hline \makecell{ CAN Intrusion \\ Dataset(OTIDS)} \cite{OTIDS[24]} & Real & Real;Dos;Fuzzing& - \\
    Survival Analysis Dataset\cite{SurvivalADataset} & Real &                   \makecell{Real;Dos;Fuzzing;\\targeted ID}&  - \\
     Car Hacking Dataset\cite{CarHackingDataset1,CarHackingDataset2} & Real &                   \makecell{Real;Dos;Fuzzing;\\targeted ID}& -  \\
      SynCAN Dataset\cite{CANet}& Simulated &                           \makecell{ Simulated;\\targeted ID} & Simulated  \\
        \makecell{ Automotive CAN Bus  \\Intrusion Dataset v2\cite{Auto.CANDatasetV2}  }& Real &\makecell{ Simulated;\\targeted ID} & Simulated \\
        Can Log Infector & Real & -&   \makecell{ Provide 7 types \\codes to simulate}  \\
        ROAD Dataset \cite{ROADDataset} & Real &          \makecell{Real;Fuzzing;\\targeted ID} &  Simulated   \\
        \hline
  \end{tabular}}
\end{table}
%
It produced by the Hacking and Countermeasure Research Lab (HCRL) of Korea University, records CAN traffic through the OBD-II port of a real vehicle \cite{CarHackingDataset1}. It includes one normal dataset and four attack datasets: DoS, Fuzzy, spoofing GEAR, and spoofing RPM. For detection window integrity, we classify each window based on its content: a window containing both normal and attack data is categorized as an attack window, with labels for both malicious and normal frames. A window containing only normal data is classified as a normal detection window.


The CAN dataset is divided into training (80\% normal, 70\% attack), validation (20\% normal, 20\% attack), and testing (remaining real CAN messages) sets. 
This imbalanced partitioning reflects real-world CAN bus conditions under attack. The same division is consistently applied across all models for training, validation, and testing. 
 


\subsubsection{ROAD~Dataset}
It is a variety of CAN attacks collected from a passenger vehicle, we choose the following 5 advanced masquerade attacks \cite{ROADDataset} :
\begin{itemize}  
%
    \item \textit{Correlated Signal Masquerade Attack}: The vehicle receives four false wheel speed values from injected malicious CAN messages, which disables the accelerator pedal and even makes a restart of the automotive electronic and electrical systems.

    \item \textit{Max Engine Coolant Temp Masquerade Attack}: The vehicle receives an alarm of ``engine coolant too high" that may mislead drivers to operate unnecessarily, due to the malicious modification of the engine coolant signal value with the maximum (0xFF) by attackers.
  
    \item \textit{Max Speedometer Masquerade Attack}: The speedometer incorrectly displays the maximum value (0xFF) due to the injected malicious CAN messages, and may cause the driver to brake urgently.
    
    \item \textit{Reverse Light Off Masquerade Attack}: The reverse lights of the vehicle are maliciously turned off when in reverse-gear, which may have danger to pedestrians due to lack of signal light alerts.
    
    \item \textit{Reverse Light On Masquerade Attack}: The reverse lights of the vehicle are maliciously turned on when in drive-gear, which may mislead the vehicles following behind and result in improper operation.
    
\end{itemize}

The ROAD dataset provides both labeled signals translated using the CAN-D algorithm \cite{CAN-D[30]} and raw, unlabeled data. Since OEMs of passenger vehicles keep their proprietary CAN signal encodings confidential and vary them across models, the CAN-D algorithm is not always reliable for translating unknown CAN signals. The translation process only involves the payload, excluding the CAN ID and timestamps.

To address this, we compare the IDs and timestamps of the translated signals (in CSV format) with those of the raw data (in log files), considering messages with identical IDs and timestamps as corresponding messages. These are then formatted to match the structure of the Car Hacking Dataset. 

\subsection{Running Performance}\label{sec_experiment_E}

Since the running performance of intrusion detection models has a significant impact on vehicle security, this paper analyses the time and memory consumption of the proposed \textsc{StatGraph} and all the multi-classification methods with the same type as \textsc{StatGraph} in \S \ref{sec_experiment_C} to evaluate their applicability in resource-constrained in-vehicle environments. 

 
\begin{figure}[!ht]
	\centering	
    \includegraphics[width=0.99\linewidth]{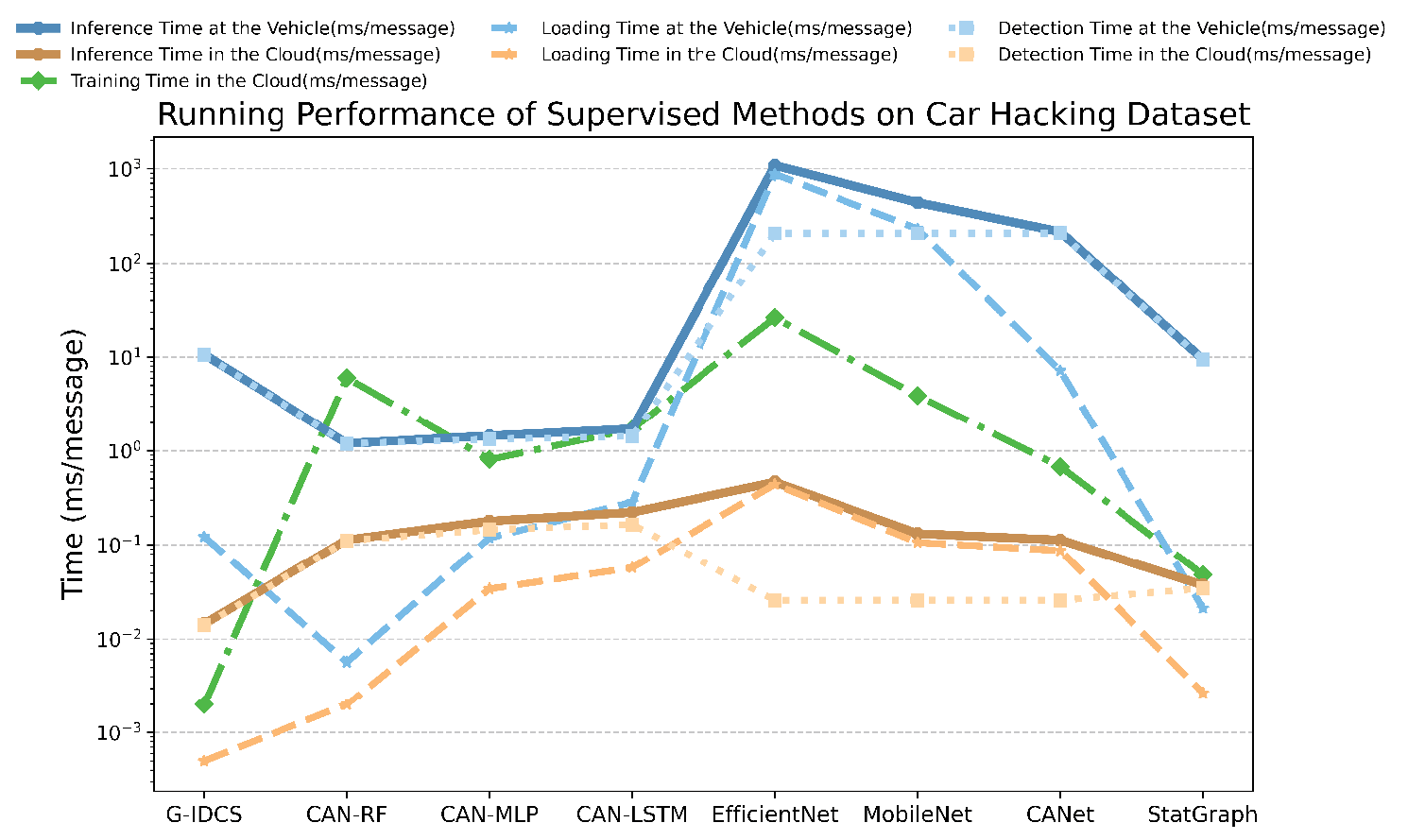}
    \includegraphics[width=0.99\linewidth]{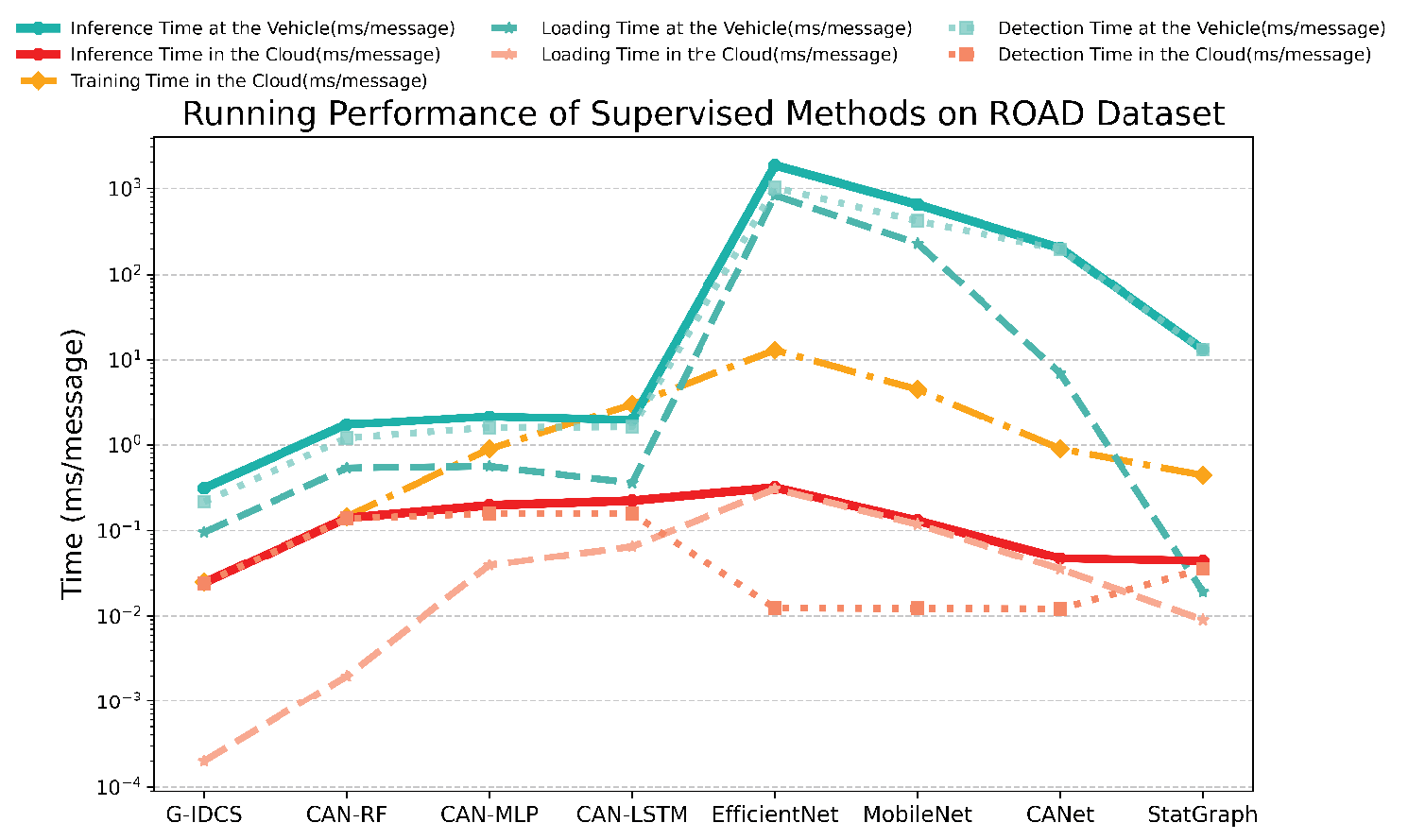}
	\caption{Running Time Comparison Evaluation.}  
	\label{runningPerformance}
\end{figure}

\begin{figure}[!ht]
	\centering	
    \includegraphics[width=0.85\linewidth]{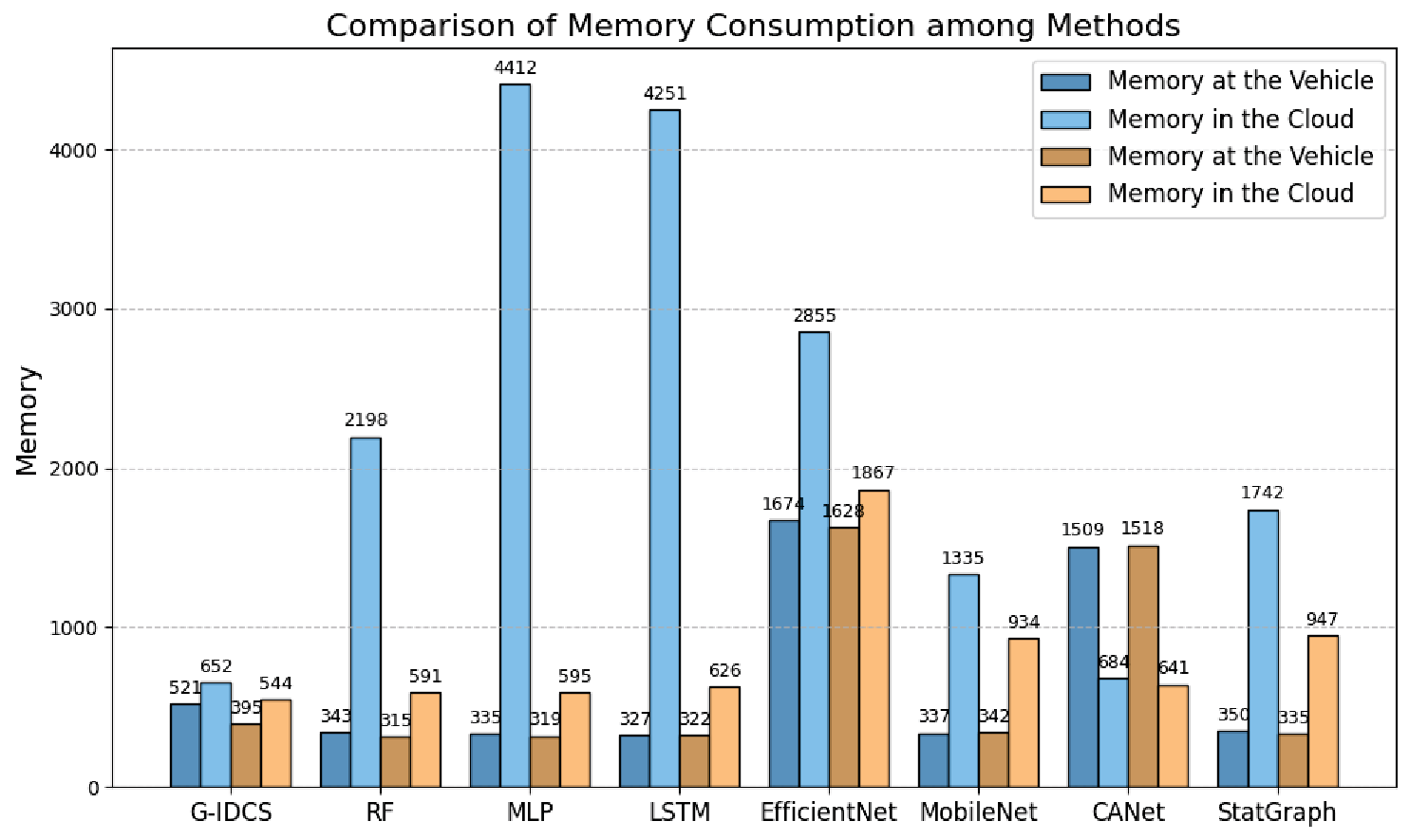}
	\caption{Memory Comparison Evaluation.}
	\label{MemoryPerformance}
\end{figure}


The experimental results, shown in Figure \ref{runningPerformance}, illustrate the average training time, loading time, testing time, and inference time consumption. The inference time is the sum of both the processing and detection times, as it includes the entire inference process, from loading the data to detection.

It is evident that cloud computing devices exhibit lower computational times (processing, detection, and inference) compared to vehicle devices across all models. This suggests that a resource-constrained environment can degrade detection speed. Notably, the G-IDCS ML\_classifier demonstrates lower time consumption in both training and inference processes, as it reduces the detection workload by using three features extracted from 200 CAN messages as inputs at once.

In contrast, the simpler CAN-RF model detects data item by item, resulting in the shortest detection time in the cloud. However, the difference is not significant on vehicle devices. Due to their complex network structures, CAN-MLP and CAN-LSTM maintain relatively stable but longer inference times, whether on vehicle or cloud devices.

EfficientNet, MobileNet, and CANet exhibit increased processing times due to the conversion of raw CAN data into image data, with CANet showing lower overall running and inference times due to its lightweight design. Finally, \textsc{StatGraph} performs well in the cloud, benefiting from its sophisticated data processing. However, its inference time increases on resource-constrained vehicle devices due to time-intensive data format processing.
Overall, the inference time trend of the methods is consistent on the two data sets. CAN-MLP, CAN-LSTM, CANet and \textsc{StatGraph} are relatively lightweight and have the potential for real-world vehicle applications.

Memory consumption results, as shown in Figure \ref{MemoryPerformance}, reveal that most models consume more memory on cloud devices than on vehicle devices, with CANet being the exception. This discrepancy contributes to faster detection speeds in the cloud, while vehicle devices offer advantages in lower power consumption and memory usage. Notably, CAN-RF, CAN-MLP, CAN-LSTM, and \textsc{StatGraph} exhibit reduced memory requirements when deployed on vehicle devices.

In summary, the inference time of IVN intrusion detection methods is heavily influenced by hardware capabilities. Although modern IVN systems have enhanced ECU computational power to manage real-time tasks \cite{50-real-time}, their processing capacity remains constrained. Consequently, instead of deploying IDS directly on ECUs, we utilize bypass monitoring devices on the CAN bus. This strategy improves intrusion detection speed while minimizing memory and energy consumption. \textsc{StatGraph} excels in real in-vehicle computing environments, meeting real-time detection requirements even under resource constraints. Furthermore, as computational power continues to advance, other models currently unsuitable for real-time detection may become viable for such applications in the future.

%
%
%
%

\begin{table*}[htb]\footnotesize
  \centering
  \caption{Test Results of each method for different attacks in Car Hacking Dataset.}\label{Result-CHD} 
  \setlength{\tabcolsep}{2.3mm}{
  \begin{tabular}{|c|c|ccccc|c|}
    \hline
     Dataset & $Attacks$  &    Methods  & Accuracy & Precision  & Recall & F1-Score & \makecell{Window Size/\\Label Number}\\
   \hline
   \multirow{20}{*}{Car Hacking Dataset}  &\multirow{3}{*}{DoS Attack} & Graph-based IDS \cite{5-20}& 0.432 & 0.766  &  0.387  & 0.514  & 1200/1 \\
				        & ~ &  G-IDCS TH\_classifier \cite{G-IDCS} & 0.3096 & 1.0000 & 0.3096& 0.4729 & 200/1 \\
                        &  &  \textsc{StatGraph} & 0.9980  &  0.9953  &  0.9721  &  \bf{ 0.9836}& 1/1 \\
                          \cline{2-8}
   & \multirow{3}{*}{Fuzzy Attack} &  Graph-based IDS \cite{5-20} & 0.458   & 0.761 &  0.426  & 0.546& 1200/1 \\
				        &  &  G-IDCS TH\_classifier \cite{G-IDCS}  &  0.3495 & 1.0000&0.3495& 0.5180& 200/1 \\
                        &  &  \textsc{StatGraph}  & 0.9970  &  0.7804  &  0.8392  &  \bf{ 0.8088} & 1/1\\ %
    \cline{2-8}
    &\multirow{3}{*}{Drive GEAR Attack} & Graph-based IDS \cite{5-20}&0.608   & 0.721 &  0.594  & 0.651 & 1200/1 \\
				       &   &  G-IDCS TH\_classifier \cite{G-IDCS} &  0.4379& 1.0000& 0.4379&0.6090 & 200/1 \\
                       &   &  \textsc{StatGraph}  & 0.9980  &  0.9996  &  0.8920  & \bf{0.9427}& 1/1\\
    \cline{2-8}
    &\multirow{3}{*}{RPM Attack} &  Graph-based IDS \cite{5-20} & 0.899   & 0.971 &  0.856  & 0.910 & 1200/1\\
				       &   &  G-IDCS TH\_classifier \cite{G-IDCS} &  0.4592 & 1.0000 & 0.4592 & 0.6294 & 200/1\\
                       &   &  \textsc{StatGraph} & 0.9980  &  0.9932  &  0.9355  &  \bf{ 0.9635}  & 1/1\\
   \cline{2-8}
   &\multirow{8}{*}{Mixing Fabrication Attack} & EfficientNet \cite{EfficientNet} & 0.0688 & 0.2526 & 0.8015 & 0.3841  & 54/1\\
                       &   &  MobileNet \cite{MobileNetV3} & 0.2815 & 0.2474 & 0.8485 & 0.3831  & 3/1\\
				       &   &  CANet \cite{CANet}& 0.9050 & 0.6447 & \bf{0.9725} &0.7754 & 54/1\\
&   &  G-IDCS ML\_classifier \cite{G-IDCS} & 0.9269 & 0.8759 & 0.8711 & 0.8735 & 200/1\\
&   &  CAN-RF \cite{114}& 0.9733&	0.9807&	0.8208	&0.8303  & 1/1\\
&   &  CAN-MLP \cite{114} & 0.9314	 & 0.9222 &	0.6731 &	0.7782   & 1/1\\
&   &  CAN-LSTM \cite{20} & 0.9415&	0.7521  &0.8672	&0.8055 & 1/1\\
                       &   &  \textsc{StatGraph} &  \bf{0.9940}  & \bf{0.9275}    &0.9533 & \bf{0.9403} & 1/1 \\
    \hline
  \end{tabular}}
\end{table*}

\subsection{Detection Performance}\label{sec_experiment_F}

The effectiveness of \textsc{StatGraph} is compared with state-of-the-art IVN intrusion detection methods, and the experimental results on the vehicle equipment show that \textsc{StatGraph} has advantages on detecting anomalies in both datasets.

Specifically, Table \ref{Result-CHD} and Table \ref{Result-ROAD} show the Accuracy, Precision, Recall and F1-score of all the methods for various attacks on Car Hacking Dataset and ROAD Dataset. As the nature of binary detection methods, Graph-based IDS \cite{5-20} and G-IDCS TH\_classifier \cite{G-IDCS} are failure to set consistent thresholds for multiple attack categories, and thus we only place the results of those multi-classification methods when detecting the Mixing Fabrication Attack and Mixing Masquerade Attack. 

On the simpler Car Hacking Dataset, \textsc{StatGraph} achieves near-perfect fine-grained detection results, with 99.40\% accuracy and 94.03\% F1-score. This is attributed to the dataset's focus on fabrication attacks, whose anomalous behavior, such as periodicity, can be easily captured by hidden rules. In contrast, the ROAD Dataset presents a greater challenge, as evidenced by the significantly lower average accuracy, precision, recall, and F1-score across most methods. This is due to the advanced masquerade attacks in the ROAD Dataset, which involve the complete replacement of legitimate messages while replicating their statistical behavior patterns, with only subtle payload threshold alterations, making them highly stealthy. Despite this, \textsc{StatGraph} remains the most advanced detector, achieving 97.91\% accuracy and 97.46\% F1-score.

The generalizability of models varies significantly. For instance, EfficientNet and MobileNetV3 perform well on the ROAD Dataset but poorly on the Car Hacking Dataset, while CAN-RF, CAN-MLP, CAN-LSTM, and CANet exhibit the opposite trend. \textsc{StatGraph} consistently achieves the highest accuracy, F1-score, and IG values across both datasets, demonstrating its superior adaptability and robustness.

\begin{table*}[htb]\footnotesize
  \centering
  \caption{Best Test Results of each method for different attacks in ROAD Dataset.}\label{Result-ROAD} 
  \setlength{\tabcolsep}{2.4mm}{
  \begin{tabular}{|c|c|ccccc|c|}
    \hline
    Dataset & $Attacks$  &    Methods  & Accuracy & Precision  & Recall & F1-Score & \makecell{Window Size/\\Label Number} \\ 
   \hline
     \multirow{23}{*}{ROAD Dataset} &\multirow{3}{*}{\makecell{Correlated signal\\ Masquerade Attack }}  &  Graph-based IDS \cite{5-20} &  0.540  &0.450 &  0.818  & 0.581 & 1200/1\\
                       & ~ &  G-IDCS TH\_classifier \cite{G-IDCS} & 0.6922 & 1.0000 & 0.6922 & 0.8181 & 200/1 \\
                       &   &  \textsc{StatGraph} & 0.9990 & 0.9944 &  0.9976 & \bf{0.9960} & 1/1\\ 
   \cline{2-8}
   &\multirow{3}{*}{\makecell{Max Speedometer\\ Masquerade Attack  }}  &   Graph-based IDS \cite{5-20} & 0.625  &   0.143  &   0.556  &  0.227& 1200/1\\
                       & ~ &  G-IDCS TH\_classifier \cite{G-IDCS} & 0.3806 & 1.0000 & 0.3806 & 0.5513 & 200/1 \\
                       &   &  \textsc{StatGraph} & 0.9618 & 0.9627 &  0.9618  & \bf{0.9622} & 1/1\\
   \cline{2-8}
   &\multirow{3}{*}{\makecell{Reverse Light Off\\ Masquerade Attack}}  &  Graph-based IDS \cite{5-20} & 0.603  &   0.448  &   0.591  &0.510 & 1200/1\\
                       & ~ &  G-IDCS TH\_classifier \cite{G-IDCS} & 0.4455 & 1.0000 & 0.4455 & 0.6164  & 200/1 \\  
                       &   &  \textsc{StatGraph} & 0.9551 & 0.9122 & 0.9551 & \bf{0.9332} & 1/1 \\
   \cline{2-8}
   &\multirow{3}{*}{\makecell{Reverse Light On\\ Masquerade Attack}}  & Graph-based IDS \cite{5-20}& 0.528  &   0.215  &   0.469  &0.295& 1200/1\\
                      & ~ &  G-IDCS TH\_classifier \cite{G-IDCS} & 0.4512 & 1.0000 & 0.4512 & 0.6218   & 200/1 \\
                       &   &  \textsc{StatGraph} & 0.9553 & 0.9132 & 0.9553 & \bf{0.9338} &1/1 \\
   \cline{2-8}
   &\multirow{3}{*}{\makecell{Max Engine Coolant\\ Temp Masquerade Attack  }}  &  Graph-based IDS \cite{5-20} & 0.830  &   0.111  &   1.000  &0.200 & 1200/1\\
                      & ~ &  G-IDCS TH\_classifier \cite{G-IDCS} & 0.1333 & 1.0000 & 0.1333 & 0.2353 & 200/1 \\
                       &   &  \textsc{StatGraph} & 0.9955 & 0.9910  & 0.9955 & \bf{0.9933} & 1/1 \\
   \cline{2-8}
   &\multirow{8}{*}{Mixing Masquerade Attack} &  EfficientNet \cite{EfficientNet}& 0.9699 & 0.5281 & 0.5232 & 0.5236 & 3/1 \\
                       &   &  MobileNet \cite{MobileNetV3} & 0.9637 &  0.5228 & 0.5084 & 0.5132& 3/1\\
				       &   &  CANet \cite{CANet}& 0.9459 & 0.4111 & 0.3292 & 0.3585& 3/1 \\
            &   &  G-IDCS ML\_classifier \cite{G-IDCS} & 0.7281 &  0.3115 & 0.2672 & 0.2811 & 200/1\\
            &   &  CAN-RF \cite{114} & 0.9699 & 0.4141 & 0.3429 & 0.3752 & 1/1\\
            &   &  CAN-MLP \cite{114} & 0.8341&0.6970&	0.4181	&	0.5227   & 1/1\\
            &   &  CAN-LSTM \cite{20} & 0.6741 & 0.9577 & 0.6354 & 0.7639  & 1/1\\
                       &   &  \textsc{StatGraph} &\bf{ 0.9897 } & \bf{0.9818} & \bf{0.9897} & \bf{0.9858} & 1/1\\ 
   \hline   
  \end{tabular}}
\end{table*}

\begin{table}[htb]\scriptsize
  \centering
  \caption{Test results of each method in mix attacks\\ (real-time detection: $N=3$, semi-real-time detection: $N=27$ and offline detection: $N=54$).}\label{ResultIG} 
  \setlength{\tabcolsep}{1.5mm}{
  \begin{tabular}{cccccccc}
    \hline
   $Method$  & Window Size & \makecell{ F1-Score \\(Car Hacking) }  & \makecell{ IG\\(Car Hacking)} & \makecell{ F1-Score \\(ROAD) } & \makecell{ IG\\(ROAD)}\\ %
   \hline  \multirow{3}{*}{EfficientNet}   & $N$ = 54 & $\underline{0.3841}$ & $\underline{0.0128}$ & 0.5236 & 0.4664 \\ 
                                           & $N$ = 27 & 0.1382  & 0.0054 & 0.3348 & 0.4957 \\
                                           & $N$ = 3 & 0.1215  & 0.0041 &  $\underline{0.5236}$  & $\underline{0.9811}$  \\
    \hline  \multirow{3}{*}{ MobileNetV3 }   & $N$ = 54 & 0.1018 &0.0046  & 0.2861 &0.4384   \\ 
                                           & $N$ = 27 & 0.1109 & 0.0055 & 0.3411 & 0.4657 \\
                                           & $N$ = 3 & $\underline{0.3831}$ &$\underline{0.2815}$ & $\underline{ 0.5132 }$ & $\underline{0.9285 }$ \\
    \hline  \multirow{3}{*}{ CANet }   & $N$ = 54 & $\underline{ 0.7754}$ & $\underline{0.8996}$ &0.1060 & 0.4664 \\ 
                                           & $N$ = 27 & 0.5755 & 0.7910 &  0.1100 & 0.4942 \\
                                           & $N$ = 3 & 0.7023  & 0.7229 & $\underline{ 0.3585}$ &  $\underline{0.9261}$\\
    \hline  \multirow{3}{*}{\textsc{StatGraph} }   & $N$ = 54 & \bf{0.9403} & \bf{0.9940}   & \bf{0.9858}  & \bf{0.9897}  \\
                                           & $N$ = 27 & \bf{0.9403} & \bf{0.9940} & \bf{0.9858}  & \bf{0.9897}   \\
                                           & $N$ = 3 & \bf{0.9403}  & \bf{0.9940}  & \bf{0.9858}  & \bf{0.9897}  \\
    \hline
  \end{tabular}}
\end{table}

\subsection{Fine-grained Potential Exploration}\label{sec_experiment_G}%
Previous studies typically labeled an entire segment of consecutive CAN messages containing at least one injection attack as an attack sample, while segments without injections were labeled as normal samples \cite{wangzi2aiai}. However, except for graph-based methods \cite{5-20, G-IDCS}, no scientific rationale has been provided for the sample size used in prior research.

To address this, we establish criteria for sample size based on the CAN bus message transmission speed and latency specifications, referencing \textit{Intelligent Connected Vehicle} and \textit{Industrial Control Internet} scenarios. The industrial automation and remote driving typically require latencies below $10$\textit{ms}. Given the CAN bus speed of $125 kbps$ - $1 Mbps$ and a maximum extended frame length of 150 bits, we calculate that the CAN bus transmits a minimum of 8 messages and an average of 34 messages every 10 ms at $500 kbps$. 
Based on these values, we define three recognition granularity intervals: $0\sim8$, $8\sim34$, $34\sim+\infty$, to evaluate the performance of coarse-grained methods at different granularity levels. Accordingly, we set detection window sizes to 3, 27 and 54 for comparative analysis. \textsc{StatGraph} is evaluated against previous methods in terms of F1-Score and IG across these intervals, with results presented in Table \ref{ResultIG}. Notably, \textsc{StatGraph} operates at a per-message detection granularity regardless of the window size $N=(3,~27,~54) $, yielding consistent results across all settings.

It is evident that \textsc{StatGraph} achieves optimal F1-Score and IG values on both datasets. On the Car Hacking Dataset, among the remaining models, CANet demonstrates the best detection performance when $N=54$, indicating that CANet requires a larger detection window to accumulate sufficient data for effective characterization. Furthermore, on the ROAD Dataset, where normal data constitutes over 90\% of the dataset, misleading detection models tend to classify most data as normal. Although the other three methods exhibit competitive IG values, their F1 scores consistently fall short of \textsc{StatGraph}, underscoring its superior detection accuracy and robustness in imbalanced data scenarios.



\subsection{Parameters Analysis}\label{sec_experiment_H}

\begin{figure}[!ht]
	\centering	
	\includegraphics[width=0.8\linewidth]{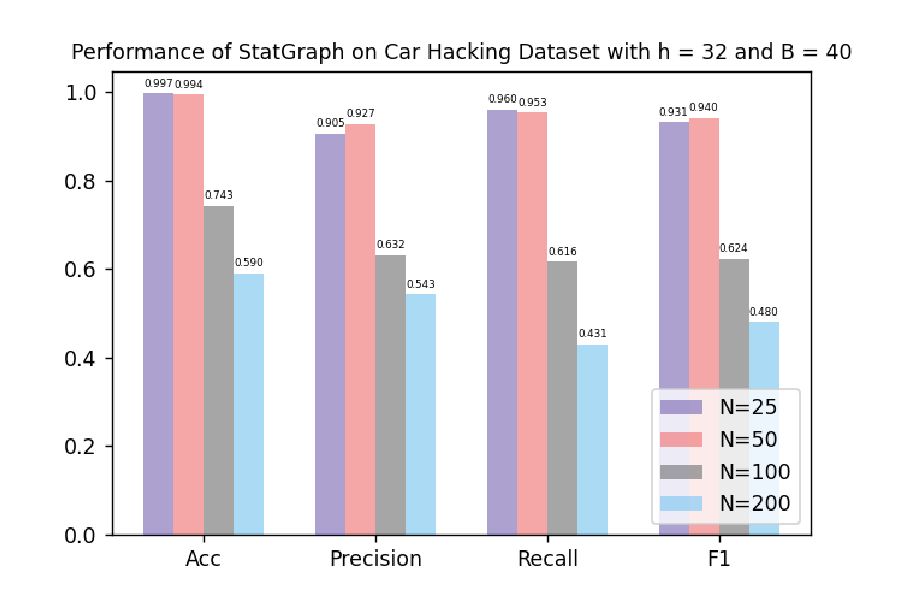}
    \includegraphics[width=0.8\linewidth]{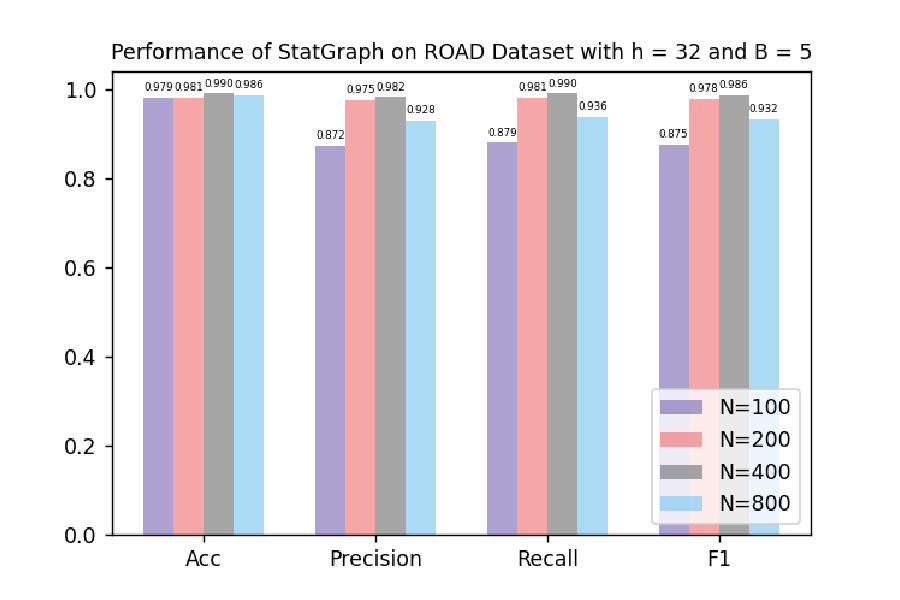} 
	\caption{Performance of \textsc{StatGraph} in Different Detection Window Size $N$.}
	\label{figsensitiveN}
\end{figure}

We evaluate the effects of different parameters that can impact the performance of \textsc{StatGraph} to find the optimal detection window size $N$, number of hidden layers units $h$ and batchsize $B$. The experiments are conducted as follows:

\textit{1) Sensitivity to detection window size:} Figure \ref{figsensitiveN} illustrates the performance comparison for detection window sizes $N$ ranging from 25 to 200 (Car Hacking) and 100 to 800 (ROAD). The results indicate that the detection window size significantly influences intrusion detection accuracy. A large $N$ delays intrusion detection due to the need for more observations, while a small $N$ fails to capture sufficient contextual features. The highest F1-Score is achieved with $N=50$ for Car Hacking and $N=400$ for ROAD, respectively.

\begin{figure}[!ht]
	\centering	
	\includegraphics[width=0.8\linewidth]{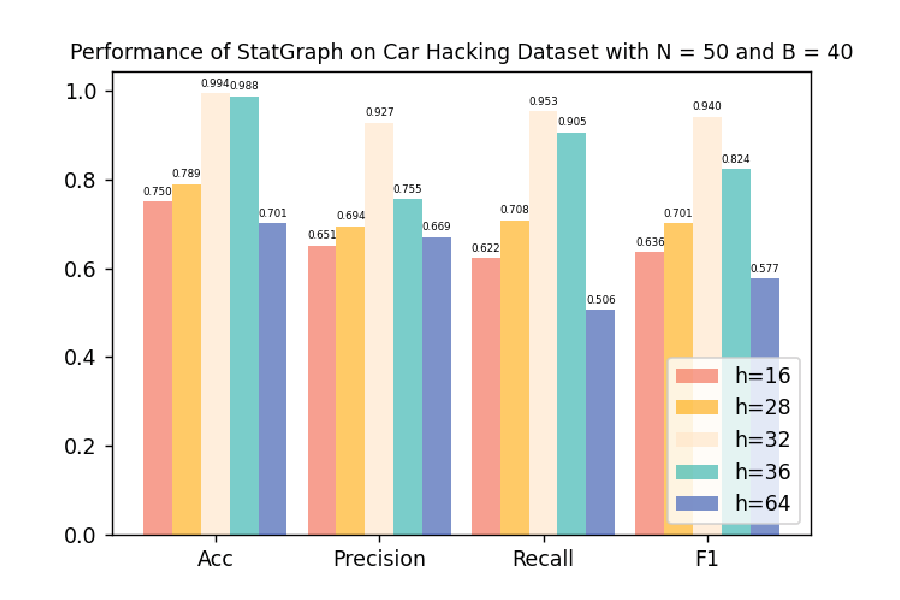} 
    \includegraphics[width=0.8\linewidth]{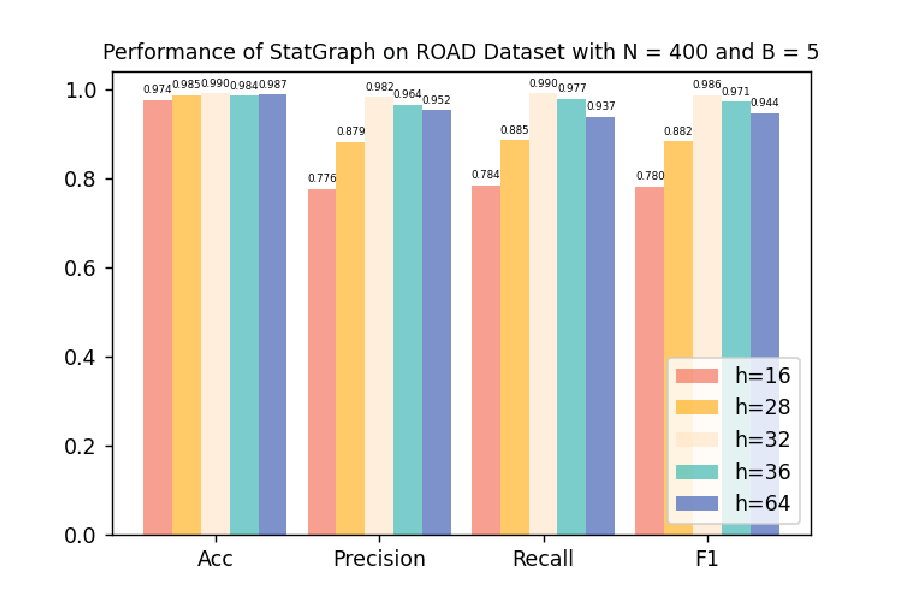}
	\caption{Performance of \textsc{StatGraph} in Different Hidden-layer Units $h$.}
	\label{figsensitiveh}
\end{figure}

\textit{2) Sensitivity to number of hidden layer units $h$:} Figure \ref{figsensitiveh} compares performance across different latent space dimensions, ranging from 16 to 64. The results reveal that excessively large dimensions lead to a significant decline in F1-Score, likely due to overfitting. Conversely, overly small dimensions result in substantial information loss during encoding, severely degrading GCN performance. A latent space dimension of $32$ achieves the best F1-Score on both datasets, enabling \textsc{StatGraph} to balance training efficiency and accuracy.

\begin{figure}[!ht]
	\centering	
    \includegraphics[width=0.8\linewidth]{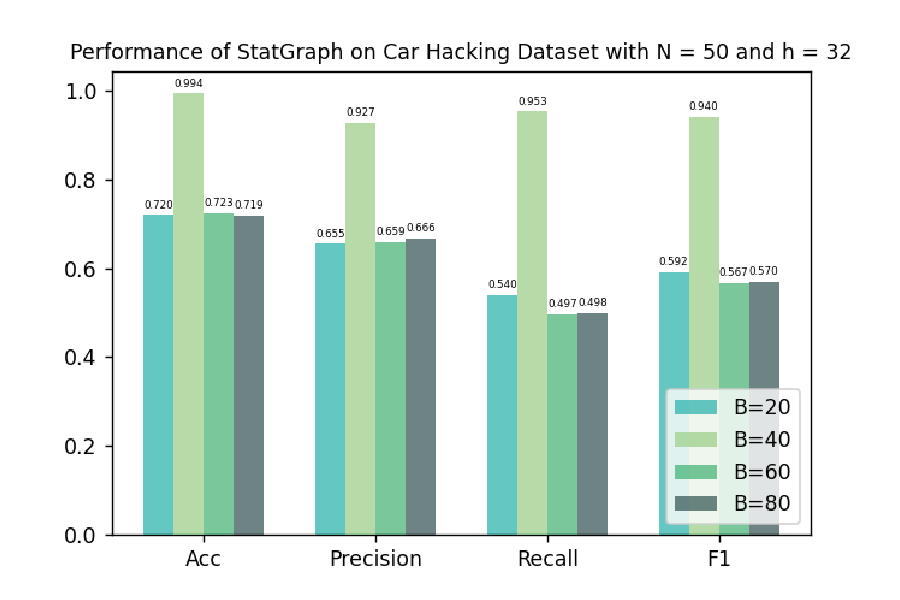} 
    \includegraphics[width=0.8\linewidth]{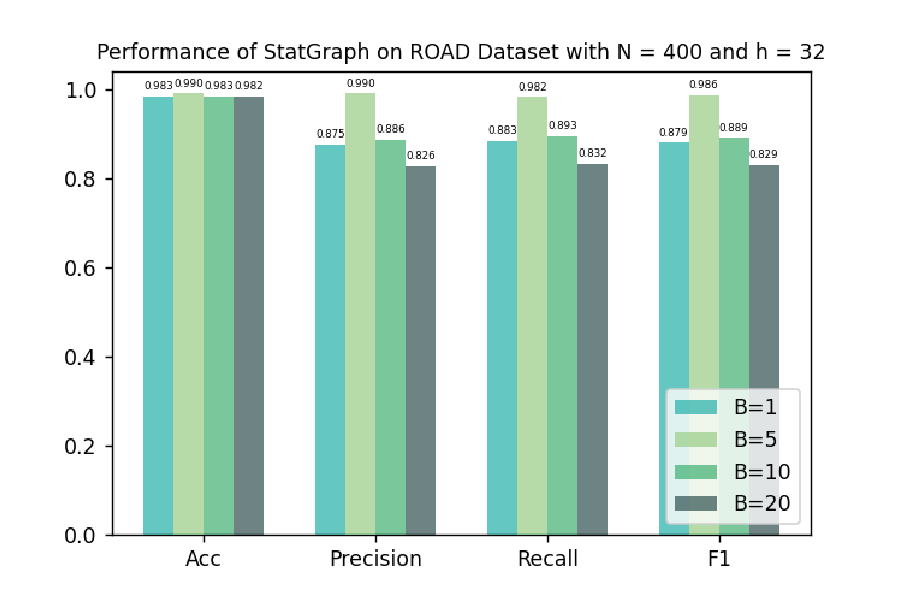}
	\caption{Performance of \textsc{StatGraph} in Different Batchsize $B$.}
	\label{figsensitive batchsize}
\end{figure}

\textit{3) Sensitivity to batch size:} Figure \ref{figsensitive batchsize} compares performance for batch sizes ranging from (20, 80) on the Car Hacking dataset and (1, 20) on the ROAD dataset. The results indicate that both excessively large and small batch sizes degrade intrusion detection performance. The highest F1-Score is achieved with a batch size $B$ of $50$ for the Car Hacking dataset and $5$ for the ROAD dataset, respectively.
%
%

\section{Discussion}\label{section6}
The experimental results demonstrate that \textsc{StatGraph} consistently outperforms other methods on both the Car Hacking Dataset and the ROAD dataset when evaluated individually. While several alternative approaches also achieve acceptable performance—such as EfficientNet and MobileNetV3 on the ROAD dataset, and CANet, G-IDCS ML\_classifier, CAN-RF, CAN-MLP, and CAN-LSTM on the Car Hacking dataset—\textsc{StatGraph} excels when assessed across both datasets simultaneously. This underscores its superior generalization capability. The transformation of multivariate time series data from IVN CAN messages into graph-structured representations facilitates a deeper understanding of the data's intrinsic features. Furthermore, leveraging multi-view graphs and selecting optimal perspectives enhance the model's generalization ability and refine its classification granularity.

Achieving fine-grained intrusion detection, where each individual message is identified, could enable attack source localization. However, the challenge of effective traceability and defense in IVN remains an unresolved issue. Future research should focus on designing novel graph views that closely align with the interaction characteristics of each ECU. Additionally, incorporating a traceability mapping module for linking messages to their source ECUs in the IVN could enhance the effectiveness of intrusion detection systems.

It is important to highlight that the ROAD dataset makes significant contributions by providing diverse samples of advanced masquerade attacks. However, the failure of many intrusion detection methods on the ROAD dataset can be attributed to the stealthy nature of masquerade attacks, which replace legitimate messages entirely while mimicking their statistical patterns, with only the payload thresholds altered.

Furthermore, a common limitation of supervised learning models is their reliance on labeled data, which hinders their ability to detect unknown attacks. As such, self-supervised and unsupervised learning approaches represent promising avenues for future research in intrusion detection.

Another challenge is the issue of imbalanced datasets, which can lead to suboptimal model performance. This can be addressed by employing data generation methods to mitigate the problems of sample imbalance and insufficient attack samples.

\section{Conclusion}\label{section7}
Intrusion detection using deep neural networks and graph learning technologies has opened new avenues for enhancing the security of IVN. However, improving detection granularity and recognizing complex attacks remains a significant challenge. 
To address this, we propose \textsc{StatGraph}, which achieves effective and fine-grained IVN intrusion detection, covering both simple (e.g., DoS) and sophisticated (e.g., masquerade) attacks through multi-view statistical graph learning on CAN messages, capturing variations in periodicity, payload, and signal combinations. Additionally, we introduce the concept of Identification Granularity (IG) to quantify detection granularity, guiding future advancements. Extensive experiments show that \textsc{StatGraph} outperforms traditional ML-based methods, deep neural network models, and graph-based methods in runtime performance, detection accuracy, and fine-grained evaluation on both the Car Hacking and ROAD datasets. 
In future work, we will explore attack traceability based on fine-grained detection and unsupervised intrusion detection mechanisms for highly imbalanced datasets.

\vspace{-25pt}
\begin{IEEEbiography}[{\includegraphics[width=1in,height=1.25in,clip,keepaspectratio]{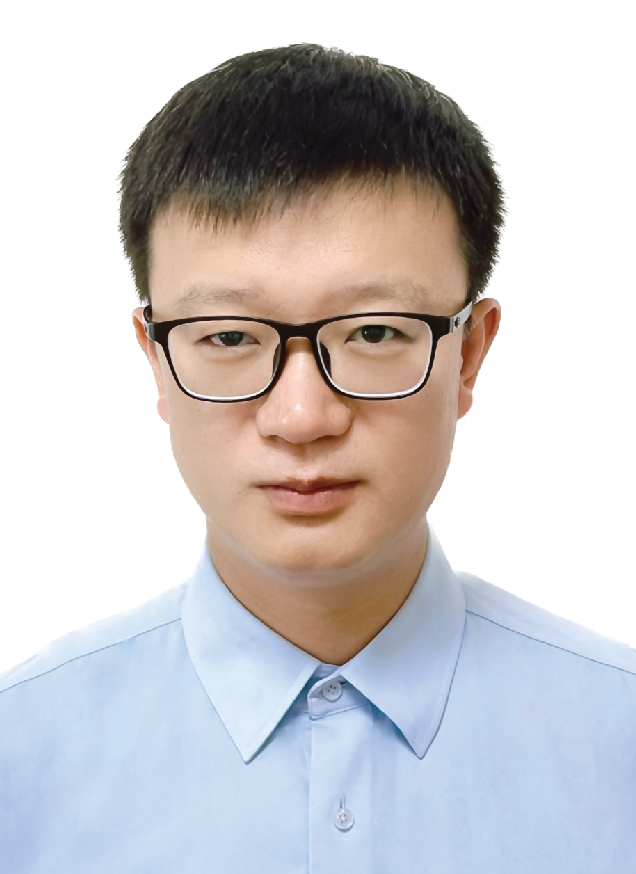}}]{Kai Wang}(Member, IEEE) is currently a professor with the School of Computer Science and Technology, Harbin Institute of Technology (HIT), Weihai. He received the B.S. and Ph.D. degrees from Beijing Jiaotong University. He has published more than 40 papers on IEEE TITS, IEEE TCE, ACM TOIT, ACM TIST, etc. His current research interests include applied machine learning for network attack detection and information forensics. He is a Member of the IEEE and ACM, and a Senior Member of the China Computer Federation (CCF).
\end{IEEEbiography}

\vspace{-25pt}
\begin{IEEEbiography}[{\includegraphics[width=1in,height=1.25in,clip,keepaspectratio]{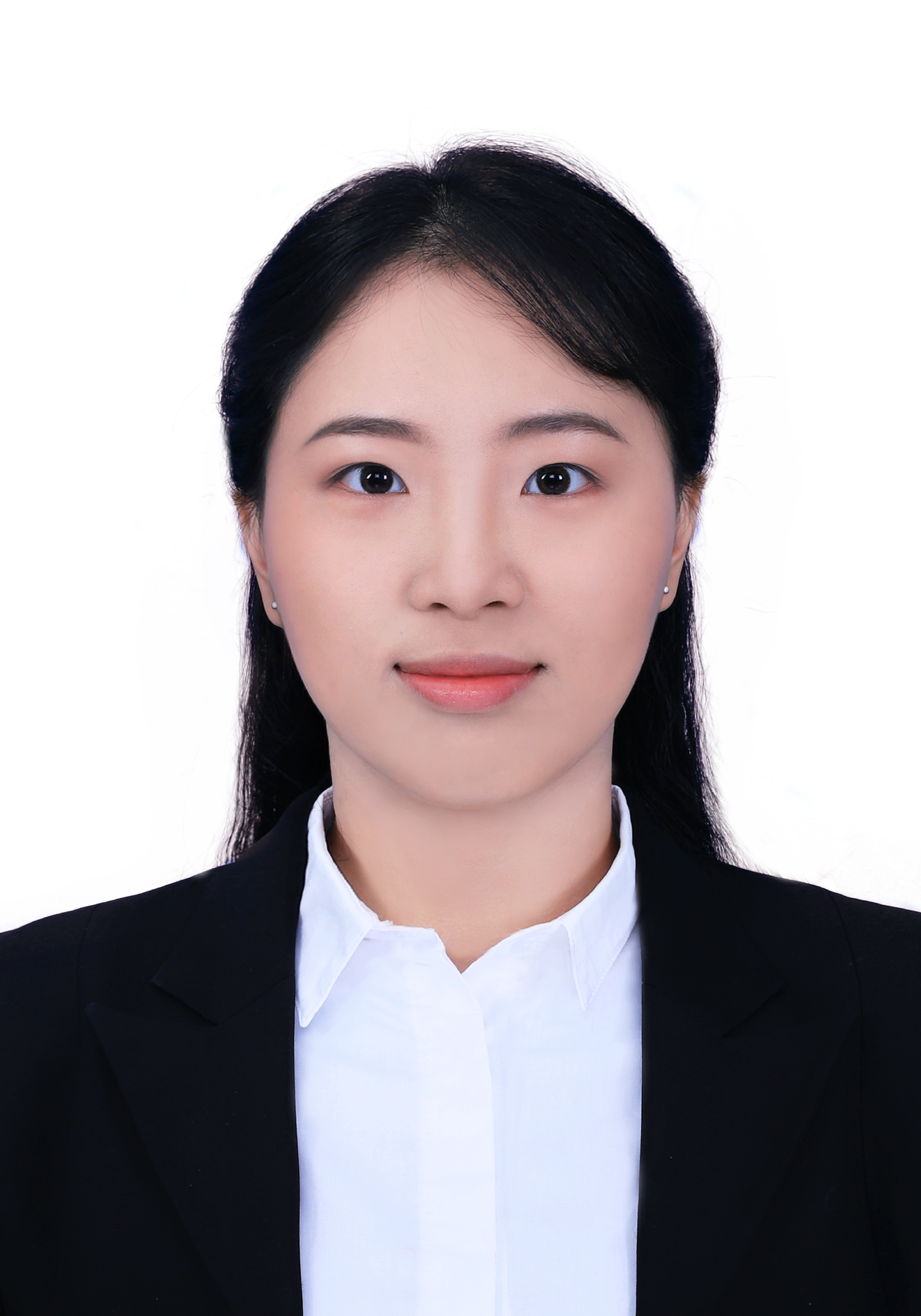}}]{Qiguang Jiang}
received the B.S. degree in Information and Computing Science from the Harbin Institute of Technology (HIT), Weihai, China. She is currently pursuing the master's degree in computer science and technology with the Harbin Institute of Technology (HIT), China. Her research interests include intelligent and efficient in-vehicle intrusion detection models.
\end{IEEEbiography}

\vspace{-20pt}
\begin{IEEEbiography}[{\includegraphics[width=1in,height=1.25in,clip,keepaspectratio]{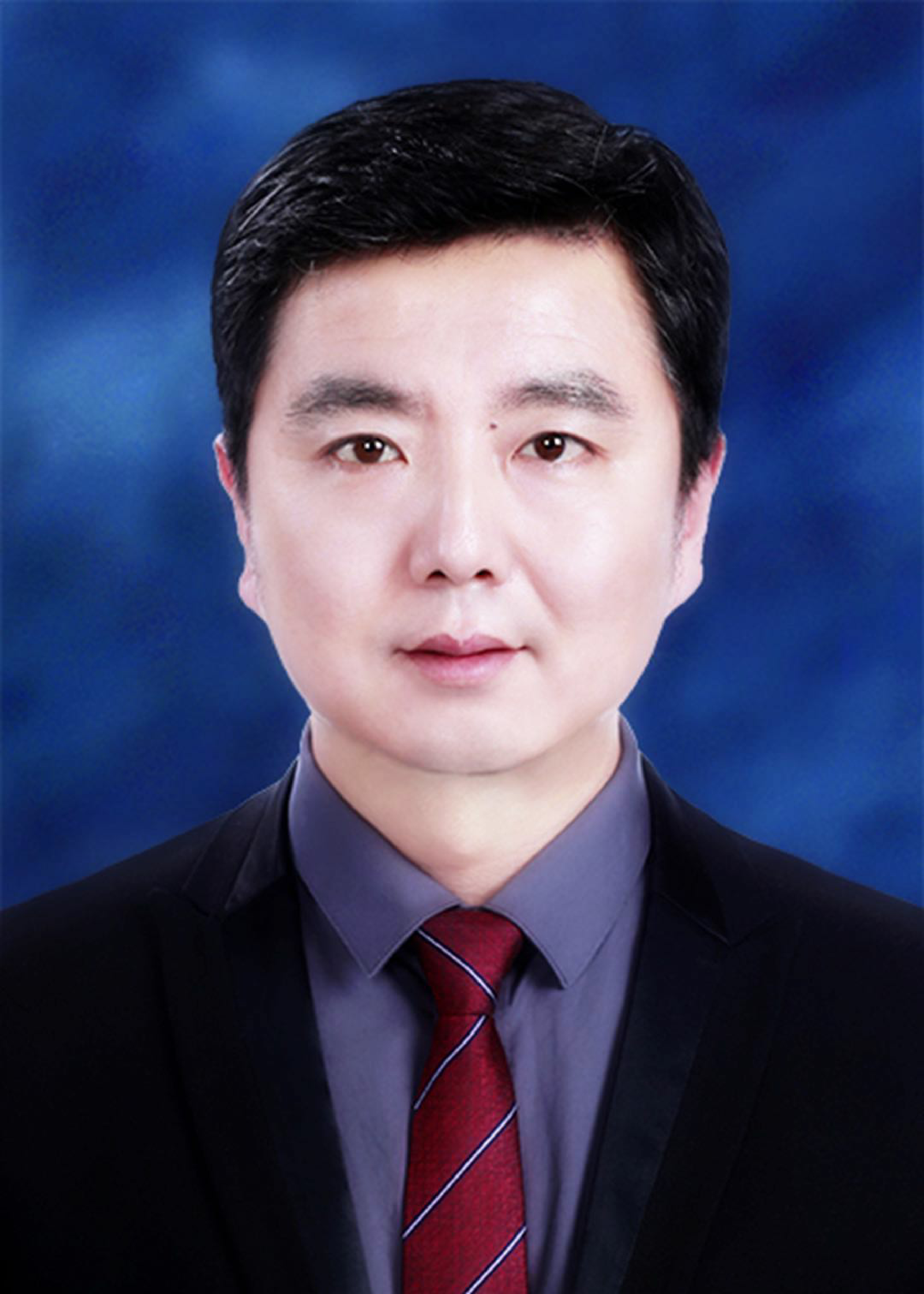}}]{Bailing Wang}
received the Ph.D. degree from the School of Computer Science and Technology, Harbin Institute of Technology (HIT), China, in 2006. He is currently a Professor with the Faculty of Computing, HIT. He has published more than 80 papers in prestigious international journals and conferences, and has been selected for the China national talent plan. His research interests include information content security, industrial control network security, and V2X security.
\end{IEEEbiography}

\vspace{-25pt}
\begin{IEEEbiography}[{\includegraphics[width=1in,height=1.25in,clip,keepaspectratio]{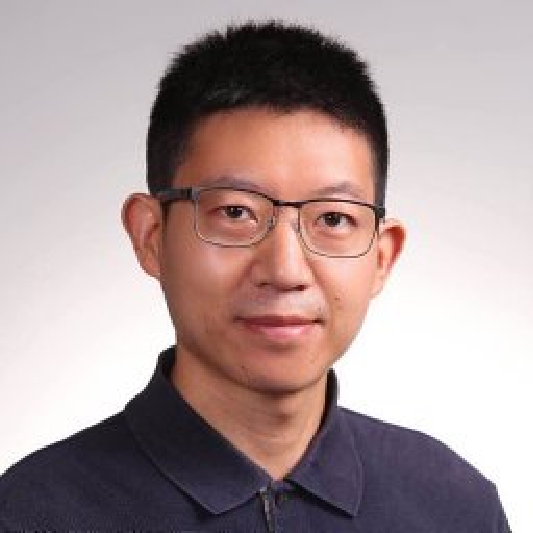}}]{Yulei Wu} (Senior Member, IEEE) is an Associate Professor with the Faculty of Engineering and the Bristol Digital Futures Institute at University of Bristol, UK. He received his Ph.D. degree in Computing and Mathematics and B.Sc. (1st Class Hons.) degree in Computer Science from the University of Bradford, United Kingdom. His research mainly focuses on network digital twins, native AI networks and systems, edge AI, and trustworthy AI. He has published over 10 authored/edited monograph books, and over 150 peer-reviewed research papers in prestigious international journals and conferences. Dr. Wu serves as an Associate Editor of IEEE TNSM and IEEE TNSE, as well as an Editorial Board Member of Computer Networks, Future Generation Computer Systems, and Nature Scientific Reports at Nature Portfolio. He is chairing an IEEE Special Interest Group (SIG) on Ethical AI for Future Networks and Digital Infrastructure. He is a Senior Member of the ACM.
\end{IEEEbiography}
\vspace{-25pt}
\begin{IEEEbiography}[{\includegraphics[width=1in,height=1.25in,clip,keepaspectratio]{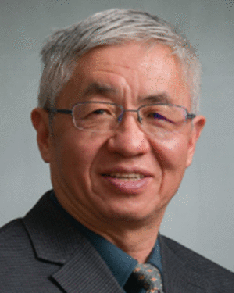}}]{Hongke Zhang}(Fellow, IEEE) is currently a Professor with the School of Electronic and Information Engineering, Beijing Jiaotong University, Beijing, China, where he currently directs the National Engineering Center of China on Mobile Specialized Network. He received the Ph.D. degree in communication and information system from the University of Electronic Science and Technology of China, Chengdu, China, in 1992. His current research interests include architecture and protocol design for the future Internet and specialized networks. He currently serves as an Associate Editor for the IEEE TNSM and IEEE Internet of Things Journal. He is an Academician of China Engineering Academy.
\end{IEEEbiography}


\vfill

\end{document}